\documentclass[final,5p]{elsarticle}

\usepackage{amsmath,graphicx}

\usepackage[mathcal]{euscript}
\usepackage{amssymb,amsmath}
\usepackage{algpseudocode}
\usepackage[]{algorithm2e}
\usepackage{verbatim}
\usepackage{graphicx}
\usepackage{longtable}
\usepackage{rotating}

\usepackage{bbm}
\usepackage{dsfont}

\usepackage{multirow}
\usepackage{tabularx}
\usepackage{bigints}
\usepackage{tikz}
\usepackage{multirow} 
\usetikzlibrary{arrows,backgrounds,snakes}
\usepackage{amsmath,amstext,amsfonts,amssymb}
\usepackage{amsbsy}
\usepackage{amsthm}
\usepackage{diagbox}
\usepackage{mathabx}
\usepackage{color}
\usepackage{morefloats}
\usepackage{float}
\usetikzlibrary{arrows,shapes,positioning,shadows,trees}
\usetikzlibrary{decorations.pathreplacing}

\usepackage{caption}
\usepackage{subcaption}

\definecolor{darkred}{rgb}{0.5,0.0,0.0}
\definecolor{darkblue}{rgb}{0.2,0.1,0.6}

\newcommand{\ul}{\underline}

\newcommand{\tc}{\textcolor}

\newtheorem{Corollary}{Corollary}

\newtheorem{theorem}{Theorem}[section]

\newtheorem{proposition}[theorem]{Proposition}

\begin{document}

\begin{frontmatter}



\title{A refined consumer behavior model for energy systems: Application to the pricing and energy-efficiency problems}


\author[1]{Chao Zhang}
\author[2]{Samson Lasaulce}
\author[3]{Li Wang\corref{cor1}}
\ead{li.wang.csu@csu.edu.cn}
\author[4]{Lucas Saludjian}
\author[5]{H. Vincent Poor}

\cortext[cor1]{Corresponding author}
\address[1]{School of Computer Science and Engineering, Central South University, Changsha 410083, China}
\address[2]{CRAN (CNRS and University of Lorraine), 54000 Nancy, France}
\address[3]{School of Mathematics and Statistics, Central South University, Changsha 410083, China}
\address[4]{RTE France, 92800 Puteaux, France}
\address[5]{Princeton University, 08544 Princeton, United States}

\begin{abstract}
%To describe the interaction between an energy provider and energy consumers, in part for simplicity, concave and non-decreasing utility functions are widely used to model the behaviors of the consumers. However, this assumption may be questioned in practice e.g., because the quality of experience of the consumer does not follow such a model. Inspired by prospect theory, we therefore propose to refine the consumer's model by considering sigmoidal utility functions.  Introducing the notion of reference point, the proposed model is more general and better matches the real behavior of the consumers. Exploiting this refined model, we pose and solve the problem of maximizing the sum-utility for a relevant class of utility functions. The derivation of the best solution allows us to design a new inclining block rates (IBR) pricing policy. This policy is by construction optimal in terms of social welfare but it has also the merit of flattening the power consumption and reducing the peak power. We also show how to maximize the energy-efficiency by a low-complexity algorithm. When compared to existing policies, simulations fully support the benefit from using the proposed approach.   
\textcolor{black}{The sum-utility maximization problem is known to be important in the energy systems literature. The conventional assumption to address this problem is that the utility is concave. But for some key applications, such an assumption is not reasonable and does not reflect well the actual behavior of the consumer. To address this issue, the authors pose and address a more general optimization problem, namely by assuming the consumer's utility to be sigmoidal and in a given class of functions. The considered class of functions is very attractive for at least two reasons. First, the classical NP-hardness issue associated with sum-utility maximization is circumvented. Second, the considered class of functions encompasses well-known performance metrics used to analyze the problems of pricing and energy-efficiency. This allows one to design a new and optimal inclining block rates (IBR) pricing policy which also has the virtue of flattening the power consumption and reducing the peak power. We also show how to maximize the energy-efficiency by a low-complexity algorithm. When compared to existing policies, simulations fully support the benefit from using the proposed approach.}
\end{abstract}


%\item General framework for designing clustering methods in energy systems optimization problems.
%\item Novel clustering method (DMOC) by taking the use of clustered data in energy system optimization problems into account .
%\item Automatically extract the features relevant to the optimization problems.
%\item Application of the new framework to two important problems: the pricing problem and the power consumption scheduling.
%\item Significant performance improvement compared to existing clustering methods. 

\end{frontmatter}
\section{Introduction}
\label{sec:intro}

The smart grid concept encompasses many technological breakthroughs when compared to existing energy networks, and electricity networks in particular. Smart grid deployment will involve big changes at all levels of the electricity networks, at the generation side, at the transmission side, and at the distribution side \cite{Fang-survey-2012}. In particular, the possibility of having bidirectional information and energy flows will be an instrumental element of modern electricity networks. To be specific, demand response (DR) has been proposed to shape the demand at the consumers' side based on the received information from the energy providers \cite{Wang-survey-2017}\cite{Hui-AE-2020}. One fundamental benefit from such an approach is to be able to overcome one of the main limitations of existing power grids, namely, the peak power problem. This approach has been shown to be beneficial not only at the consumers' side, but also for flattening the overall power consumption and thus to reduce the cost incurred by the provider. Price-based DR is one of the most widely considered DR techniques, where demands are shaped through time-dependent pricing. \textcolor{black}{By implementing pricing-based demand response programs, the consumption behaviors are guided through pricing tariffs. To flatten power consumption, the peak demand results in price escalation, which can in return reduce the actual consumption.} Several different pricing strategies have been proposed in previous works, for instance, time-of-use pricing \cite{Wang-AE-2015}, critical peak pricing \cite{Liang-AE-2021}, peak-load pricing \cite{Liang-TSG-2013}, real-time pricing (RTP) \cite{Chen-ICASSP-2011}\cite{Zhang-AE-2021}. Although most of these pricing tariffs have constant flat rates which is independent of the consumption, they can be combined with inclining block rates (IBRs) \cite{Reiss-RES-2005}, for which the electricity price rises when the consumption is beyond a given threshold. The way to obtain these pricing tariffs can be categorized according to their objectives in DR problems, such as minimization of electricity cost \cite{Khayyam-JPS-2012}, maximization of sum-utility (social welfare) \cite{Shi-SGC-2016}, minimization of aggregated power consumption \cite{Jiang-EP-2011}. In this paper, we consider an optimization problem where the pricing tariff is designed to induce a consumption behavior that allows social welfare to be maximized, social welfare being usually defined as the aggregate gain brought by all the consumers minus the electricity cost. 

A first prerequisite to conducting the analysis of social welfare maximization problem is to choose a model for the behavior of the consumers. With the approach adopted in this paper, this choice amounts to defining an appropriate utility function for the consumers. In the related literature, quadratic utility functions are intensively used; in particular, they are chosen to account for what is called the satisfaction level of a consumer. By introducing satisfaction parameters in the quadratic utility function, the authors of the seminal paper \cite{MR-SGC-2010} design an RTP scheme which differentiates the energy usage in time and a level that can achieve the global optimal performance. With the same utility function, the authors of \cite{Samadi-TSG-2012} propose a Vickrey-Clarke-Groves (VCG) mechanism for demand side management to maximize social welfare. Modelling the satisfactory level as the Euclidean distance between the real consumption and the target consumption, the authors of \cite{Jiang-CDC-2011} and \cite{Jiang-Allerton-2011} provide a distributed algorithm to balance the supply and demand in the presence of random renewable energy. \textcolor{black}{Other types of utility functions have also been considered, such as the logarithm function in \cite{Fan-TSG-2012} and the linear function in \cite{Li-SGC-2011}.} These utility functions are assumed to be non-decreasing and concave, which means that the comfort obtained by the consumer will gradually get saturated when the energy consumption reaches the target level. This is a reasonable assumption when the real consumption is close to its target level. However, the concavity also indicates the marginal benefit, referred to as the additional satisfaction that a consumer receives when the additional consumption has been used, can achieve its maximum with zero consumption. Unlike some cases in economics with decreasing marginal benefit assumption, this might be in contradiction with the practical experience in power systems \cite{Good-AE-2019}, namely, what is actually experienced by the consumer. For instance, if the available power at the consumer is less than a threshold, the consumer may be unable to realize the desired task, meaning a very low quality of experience (QoE) \cite{Valentin-2015} as long as the threshold is not reached. If the consumer is an individual, a typical situation would be that an appliance cannot be put on if the available power is too small. If the consumer is a company or an aggregator, not having enough power at its disposal may prevent a service from being provided or a transaction to occur on the energy market. Assuming a concave utility appears to be non-appropriate and a more general utility function model needs to be considered and studied. This is precisely the purpose of this paper. Indeed, in contrast with the state-of-the-art literature, instead of using  concave utility functions, we propose a sigmoidal (S-shaped) utility function to better model the behaviors of consumers, especially in the low consumption power regime. Inspired by prospect theory, we introduce a new degree of freedom or parameter $\lambda$ in the utility function, which can represent the effect of utility framing \cite{PT-1979}\cite{Saad}. Regarding demand response problems, as the sigmoidal function is not convex, conventional approaches to solve corresponding optimization problems (e.g., utility maximization or cost minimization) need to be modified or reformulated. To this end, several analytical results have been found and new approaches are accordingly proposed. In addition, significant improvements have been observed in simulations when comparing the proposed schemes with existing techniques. In what follows, we describe more accurately the literature related to the analysis conducted in this paper.

%For instance, considering the comfort level associated with the use of lighting, the comfort increasing speed should be negligible before a threshold where the brightness can not support the task. When the brightness (or consumption) reaches a critical point where it is barely enough to work, the marginal benefit reached its maximum and is going to decrease until the saturated consumption. The concave function obviously violates the practical rules in low consumption levels. To this end, we propose  a novel utility function here to better model the comfort level of  consumers, especially in low consumption regime. To the best of authors' knowledge, this is the first work to mathematically model the consumers' behaviors with sigmoidal functions.

Enabling two-way communications in smart grid systems, DR is performed at the consumer side (residential district) to promote the interaction between consumers and the energy provider with the aim of not only cutting down their energy bills but also enhancing their comfort level represented by utility functions. Smart pricing tariffs have been designed by power utilities emerge as a promising technology for incentivizing consumers to reschedule their energy usage patterns. To maximize social welfare with current grid capacity, \cite{MR-SGC-2010} has demonstrated the optimality of RTP with quadratic utility functions. Even though the unit rate of electricity varies from one time-slot to another time-slot, it remains unchanged regardless of the power consumption. Alternatively, RTP combined with IBR has been proposed in \cite{MR-TSG-2010}\cite{Zhuang-TSG-2013}, either to achieve a desired trade-off between the electricity payment and the waiting time of appliances, or to reduce both the electricity cost and the peak-to-average ratio. \textcolor{black}{In our model, since IBR based pricing tariff and sigmoidal functions both have two segments (two price levels for IBR, convex part and concave part for sigmoidal functions), it could be beneficial to use pricing tariffs with IBR.} Although this pricing tariff is shown to be efficient and attracts lots of attention in different applications, its optimality has neither been formally discussed nor proved. \textcolor{black}{One of the main contributions of the present work is to mathematically prove the optimality of IBR pricing tariffs in terms of social welfare, but not only by experiments/simulations validation.}

In addition, increasing electricity prices and concerns related to greenhouse gas effects have given more momentum to the problem of designing energy-efficient systems. To enhance energy savings, information and communication technologies are playing key roles in new grid infrastructures \cite{TWC-Bu-2012}\cite{Survey-EK-2015}. As energy-efficiency problems have been widely studied in communication systems, interdisciplinary approaches can be anticipated to tackle the energy-efficiency problems in smart grids. Some well-defined metrics to assess the energy efficiency in communication systems can be applied in power systems as well, such as overall gains divided by the total consumption (see \cite{Survey-Poor-2016}\cite{Zhang-TVT-2019}), i.e., the benefit brought by unit consumption.

\textcolor{black}{The main \textbf{contributions} of this paper can be listed as follows: 1) We propose sigmoidal utility functions instead of the classical concave functions to refine the model of consumers' behavior; 2) We pose and study the associated and new social welfare optimization problem, which consists in maximizing a sum of sigmoidal functions over the unit simplex, and we provide expressions for the optimal solutions; 3) By exploiting the solutions obtained from the aforementioned optimization problem, we propose a new inclining block rates pricing tariff and prove its optimality in terms of social welfare; 4) A bisection-like algorithm is proposed to maximize energy-efficiency while ensuring the minimum requirements of each consumer.}

The structure of the paper is as follows. The model of the utility function is presented in Sec. II. To prove the optimality of our pricing tariff, in Sec. III we firstly introduce the optimal way for the provider to allocate a given consumption to consumers such that the aggregate gain of all consumers can be maximized. The IBR pricing tariff is presented in Sec. IV, along with the maximization of energy efficiency. Simulation results are provided in Sec. V.

% Below is an example of how to insert images. Delete the ``\vspace'' line,
% uncomment the preceding line ``\centerline...'' and replace ``imageX.ps''
% with a suitable PostScript file name.
% -------------------------------------------------------------------------
%\begin{figure}[htb]
%
%\begin{minipage}[b]{1.0\linewidth}
%  \centering
%  \centerline{\includegraphics[width=8.5cm]{image1}}
%%  \vspace{2.0cm}
%  \centerline{(a) Result 1}\medskip
%\end{minipage}
%%
%\begin{minipage}[b]{.48\linewidth}
%  \centering
%  \centerline{\includegraphics[width=4.0cm]{image3}}
%%  \vspace{1.5cm}
%  \centerline{(b) Results 3}\medskip
%\end{minipage}
%\hfill
%\begin{minipage}[b]{0.48\linewidth}
%  \centering
%  \centerline{\includegraphics[width=4.0cm]{image4}}
%%  \vspace{1.5cm}
%  \centerline{(c) Result 4}\medskip
%\end{minipage}
%%
%\caption{Example of placing a figure with experimental results.}
%\label{fig:res}
%%
%\end{figure}

%\vfill
%\pagebreak

\section{System model}
\label{sec:sysm}
Each consumer of the power system represents an entity which can behave independently; a consumer may represent e.g., an individual (say a subscriber), a household, a company, an energy aggregator, or a player of the energy market. The energy demand of each consumer may depend on various factors such as the time of day, climate conditions, and also the price of electricity. The energy demand also depends on the type of the users. For example, household users may have different responses to the same price than industrial sites. Different responses of end users to various pricing tariffs can be  modeled analytically by adopting the concept of utility function from microeconomics. For all users, one can define the corresponding utility function as $\widetilde{U}(x;r)$, where $x \in \mathbb{R}$ is the power consumption level of the consumer and $r \in \mathbb{R}$ is the reference point which reflects individual expectations of power consumption and may vary among users. \textcolor{black}{The reference point is proportional to the power demand and also depends on the consumer's characteristics and consumption history (see Fig.~1). A simple example could be the reference point $r$ is linearly proportional to its power demand $\omega$, i.e., $r=\alpha\omega$ where $\alpha$ represents the linear coefficient to depict the relationship between the demand and the reference point.} More precisely, for each user, the utility function represents the level of satisfaction obtained by the consumer as a function of its power consumption and the  reference point $r$. The concept of utility framing in prospect theory indicates that humans will perceive the values of a utility in terms of gains and losses based on their own reference point $r$. Being consistent with the empirical evidence, it is assumed that the utility function satisfies the following properties:

\begin{figure}[h]
   \begin{center}
        \includegraphics[width=.5\textwidth]{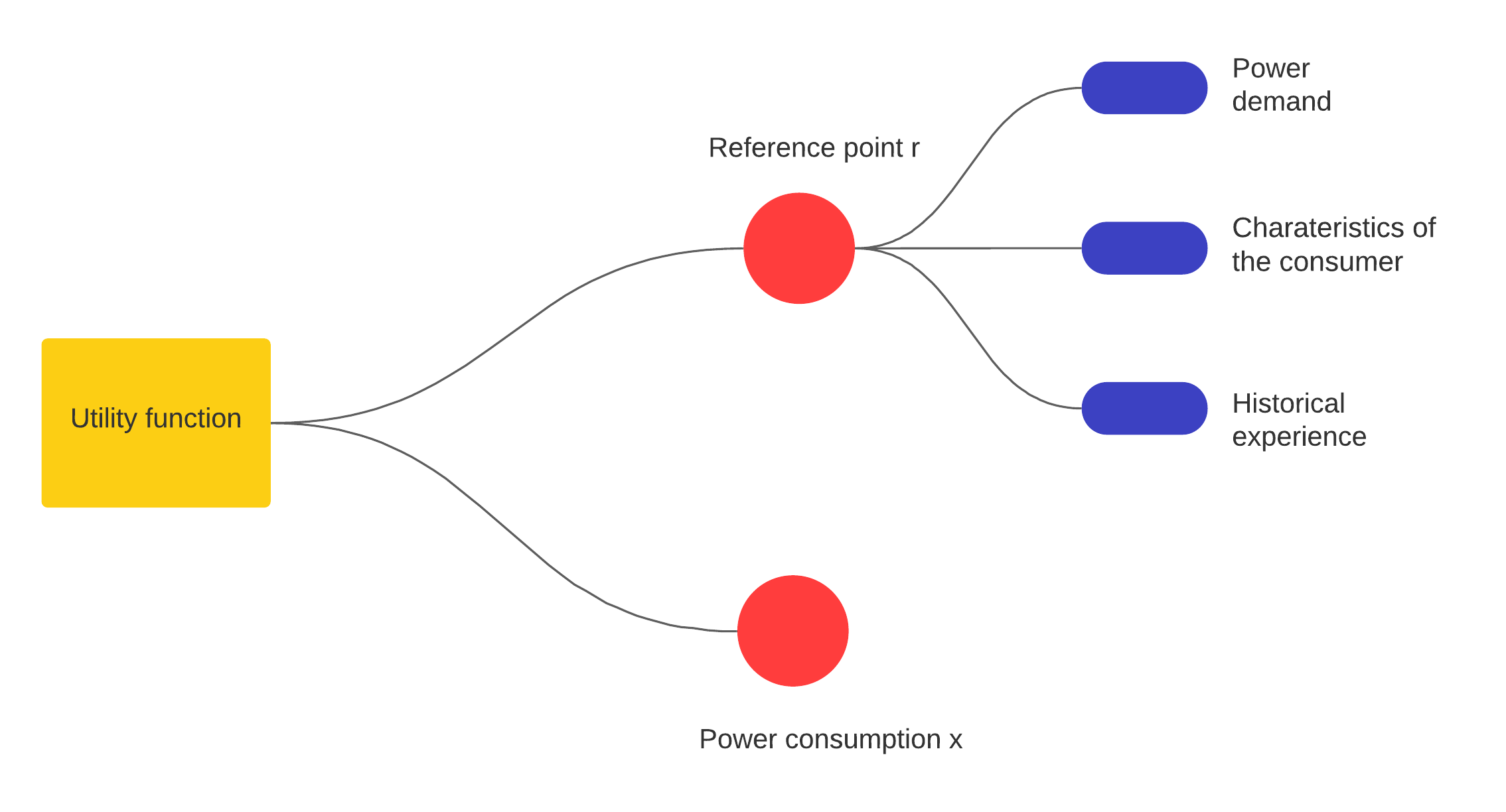}
    \end{center}
        \caption{ \textcolor{black}{Illustration of the utility function}}
\end{figure}

1) \textit{Property I:} Utility functions are non-decreasing. Consumers prefer to consume more power until the saturation power consumption is achieved. \textcolor{black}{Note that we define the utility functions for the aggregate load of different operations/tasks, rather than for the power consumption of each individual appliance. Therefore, consumers can complete more tasks if they consume more power.}  Mathematically, when $\frac{\partial \widetilde{U}(x,r)}{\partial x}$ exists, it should fulfil the following relation: 
\begin{equation}
\frac{\partial \widetilde{U}(x;r)}{\partial x}\geq 0.
\end{equation}
For notational convenience, we define
\begin{equation}
\widetilde{V}(x;r)\overset{\Delta}{=}\frac{\partial \widetilde{U}(x;r)}{\partial x}.
\end{equation}
as the marginal benefit.

2) \textit{Property II:} The marginal benefit of consumers is firstly increasing and becomes decreasing when the power consumption is beyond a threshold $\theta$. This implies that 
\begin{equation}
\forall x < \theta, \ \frac{\partial \widetilde{V}(x;r)}{\partial x}\geq 0.
\end{equation}
\begin{equation}
\forall x \geq\theta,\ \frac{\partial \widetilde{V}(x;r)}{\partial x}\leq 0.
\end{equation}
where $\theta$ is a threshold depending on $r$. The utility function is no longer supposed to be concave but appears to be sigmoidal. Unlike the utility functions proposed in \cite{MR-SGC-2010}, the maximum of marginal benefit to consumers will be achieved at a positive consumption $\theta$ rather than zero consumption. This property is in line with the motivations provided in the introduction section.

\textcolor{black}{In economics, the law of diminishing marginal utility states that, as the consumption (investment) increases, the marginal utility derived from each additional unit declines, which yields a decreasing marginal benefit. However, over recent years, researchers have found that this law is not suited to some applications in practical systems. As shown in \cite{Yang-TEM-2010}, supported by a large number of empirical evidence from a wide variety of industrial sectors, including plastics, automobile, energy, transportation, and chemicals, the relationship between investment in research and development and firm performance can be better described by sigmoidal functions. And also, the sigmoidal shape has been shown to be efficient in depicting ecological benefit functions \cite{Drechsler-EE-2020} and the alliance experience-performance relationship \cite{Tseng-CJAS-2017}. Regarding the electrical consumer utility,
empirical results (actual experience) obtained from consumers indicates that the marginal benefit at a very low consumption should be very small since a low consumption could not support the appliances to finish basic tasks (such as heating and lighting). The marginal benefit at some critical points to achieve basic goals should have a higher marginal benefit, and it should be diminishing when the power consumption continues to increase to execute optional tasks.  According to these practical experiences and the use of sigmoidal shape in  benefit-consumption relationship in the literature, we propose the sigmoidal utility functions here.}

3) \textit{Property III:} When the power consumption level is less than the reference point, it is assumed a larger $r$ yields a smaller $\widetilde{U}(x;r)$, i.e.,
\begin{equation}
\forall x<r, \  \frac{\partial \widetilde{U}(x;r)}{\partial r}\leq 0.
\end{equation}
A larger $r$ implies that the consumer is more difficult to be satisfied, and thus lead to a lower satisfactory level. \textcolor{black}{Also, the higher reference point implies a higher demand level, thus consuming the same amount of power $x$, the consumer with higher demand is less approaching to its target consumption, resulting in a lower satisfactory level.}

4) \textit{Property IV:} We assume that zero power consumption brings no benefit to the consumer, i.e., 
\begin{equation}
\widetilde{U}(x=0;r)=0.
\end{equation}
\textcolor{black}{According to these properties, the utility function should be increasing and sigmoidal, i.e., firstly convex and then concave. In addition, in a real-life setting, empirical studies have shown that decision makers tend to deviate noticeably from the rationality axioms when in the presence of uncertain reward \cite{PT-1979}\cite{Saad-2016}\cite{Tversky-1988}. Inspired by prospect theory, consumers make decisions based on the potential gain or losses relative to their specific situation (the reference point) rather than an absolute value, and feel greater aggravation for losing a certain amount of consumption than satisfaction associated with gaining the same amount of consumption (referred to as loss aversion). Very interestingly, the classical value functions used in prospect theory, such as power functions with the exponent less than one,  are increasing and sigmoidal. Therefore, adjusting classical value functions for our model, we consider the following sigmoidal utility functions:
\begin{equation}\label{eq:utility}
\widetilde{U}(x;r)=
\left\{
\begin{array}{lll}
-\lambda(r-x)^{\alpha}+\lambda r^{\alpha} & \hbox{  if \,\,\,$0\leq x< r$} \\\\
(x-r)^{\alpha}+\lambda r^{\alpha} & \hbox{ if \,\,\,$r\leq x< x_{\max}$} \\\\
(x_{\max}-r)^{\alpha}+\lambda r^{\alpha} & \hbox{ if \,\,\,$x\geq x_{\max}$} \\\\
\end{array}
\right.
\end{equation}
where choosing $0<\alpha<1$ allows one to ensure the S-shape property for the utility function, and $\lambda\geq 1$ represents a loss aversion coefficient. The reference point $r$ is typically different for each consumer and originates from its past experiences and future aspirations of profits. In our model, regardless of the common term $\lambda r^{\alpha}$ which is independent of $x$, the term $(x-r)^{\alpha}$ and $-\lambda(r-x)^{\alpha}$ can be seen as the gain and loss, respectively. 
$x_{\max}$ is considered as the saturation power of the consumer, while consuming more than $x_{\max}$ will no longer bring more benefit. The shape of the function is shown in Fig.~2.}
\begin{figure}[h]
   \begin{center}
        \includegraphics[width=.5\textwidth]{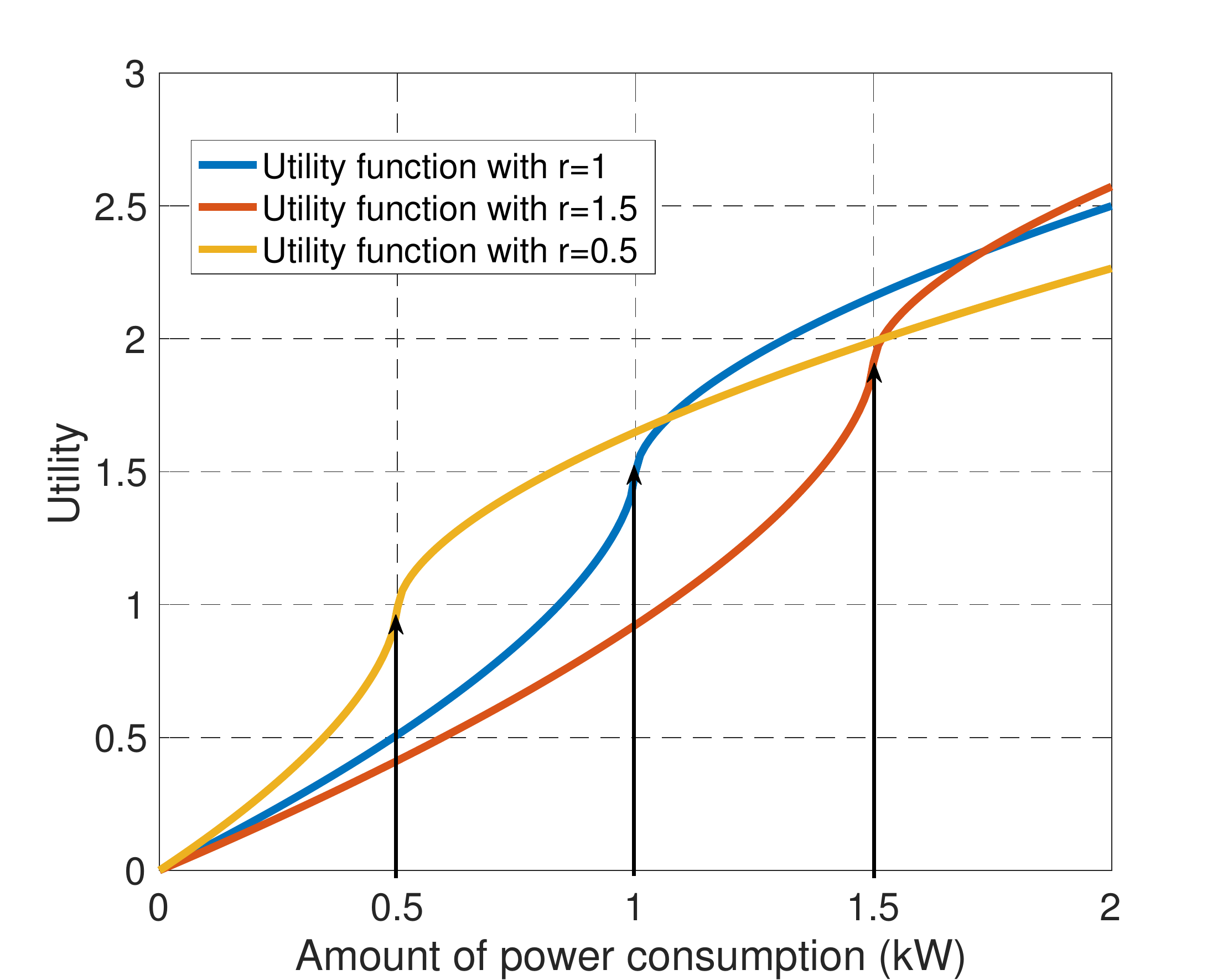}
    \end{center}
        \caption{Utility functions for the consumers by setting $\alpha=0.6$ and $\lambda=1.5$}
\end{figure}

\section{Sum-utility maximization problem}
In this paper, we consider a system consisting of one provider and $K$ consumers. Before presenting the pricing problems associated with the sigmoidal utility functions, we start by studying the allocation problem in which the provider has to allocate a fixed total power consumption  budget $\chi$ among the consumers in order for the sum-utility to be maximized. Each utility consists of a benefit function associated with consuming minus a cost induced by electricity purchase/generation. It is worth noting that the saturation part does not change the structure of the power allocation policy but only involves adding one more constraint. For the sake of clarity, we assume the maximal consumption power level $x_{\max}$ is sufficiently large and the saturation part is neglected in the rest of the paper. Hence, the utility function can be simplified as 
\begin{equation}\label{eq:utility_simplified}
U(x;r)=
\left\{
\begin{array}{lll}
-\lambda(r-x)^{\alpha}+\lambda r^{\alpha} & \hbox{  if \,\,\,$0\leq x< r$} \\\\
(x-r)^{\alpha}+\lambda r^{\alpha} & \hbox{ if \,\,\,$x\geq r$}. \\\\
\end{array}
\right.
\end{equation}
Always for the sake of clarity, the $\lambda$ and $\alpha$ are assumed to be the same for all the consumers, and thus the different consumers can be mainly distinguished by their reference points. The extension to the most general case can be conducted quite easily. By denoting $\underline{x} = (x_1,\dots,x_K)$, the sum-utility maximization problem can be written as
\begin{equation}
\begin{array}{lll}
&\underset{\underline{x}}{\max} &\sum_{i=1}^K U(x_i,r_i) \\\\
&\mathrm{s.t.} & x_i\geq 0 \\\\
&&\sum_{i=1}^K x_i\leq \chi \\\\
\end{array}
\tag{OP-G}
\end{equation}
where $x_i$ represents the power consumption of consumer $i$, $r_i$ represents the reference point of Consumer $i$. \textcolor{black}{Note that the generic problem of maximizing a sum of sigmoidal functions over a convex constraint set is a non-trivial problem; it has been shown to be NP-hard \cite{Boyd-2013}. In \cite{Boyd-2013} the authors have proposed a general algorithm to find approximate solutions only, but for some structured problems such as the problem under consideration, the problem of finding the optimal solution turns out to be tractable. For this purpose, we first introduce two subproblems and show later that the original sigmoidal programming problem OP-G can be decomposed into these two subproblems. Consequently, the solution of OP-G can be fully expressed.}

\subsection{Subproblem A}
Assume $0<a_1\leq \lambda$ and $0<\alpha<1$, the first subproblem can be written as:
\begin{equation}
\begin{array}{lll}
&\underset{x,y}{\max} &a_1x^{\alpha}+U(y;r) \\\\
&\mathrm{s.t.} & x,y\geq 0 \\\\
&&x+y\leq C_1 \\\\
\end{array}
\tag{OP-A}
\end{equation}
\textcolor{black}{where $C_1$ is a given constant.}
It is worth noting that $a_1x^{\alpha}$ is increasing w.r.t $x$ and $U(y;r)$ is increasing w.r.t $y$, and thus the inequality $x+y\leq C_1$ can be replaced by the equality $x+y=C_1$. Consequently, the first subproblem can be simplified as 
\begin{equation}
\begin{array}{lll}
&\underset{x}{\max} & a_1x^{\alpha}+U(C_1-x,r) \\\\
&\mathrm{s.t.} & 0\leq x\leq C_1.\\\\
\end{array}
\end{equation}
For notational convenience we define 
\begin{equation}
f_1(x)=a_1x^{\alpha}+U(C_1-x,r).
\end{equation}
The problem OP-A boils down to finding the maximum of the function $f_1(x)$ in the interval $[0,C_1]$. Due to the discontinuity of the derivative at $x=C_1-r$, it is difficult to express the maximum point through a single formula. However, one can check that there is at most one local maximum point for $f_1(x)$ in the interval $[0,C_1]$. By studying the first derivative of $f_1(x)$ and comparing the value of local maximum with the value at boundaries, i.e., $f_1(0)$ and $f_1(C_1)$, the solution of OP-A can be classified and written as
\begin{equation}\label{eq:solution_x_OPA}
x^{\mathrm{A}}(C_1,r,a_1)=
\left\{
\begin{array}{lll}
C_1 & \hbox{  if \,\,\,$C_1\leq r\left(\frac{\lambda}{a_1}\right)^{\frac{1}{\alpha-1}}$} \\\\
\frac{r-C_1}{(\frac{\lambda}{a_1})^{\frac{1}{1-\alpha}}-1} & \hbox{ if \,\,\,$ r\left(\frac{\lambda}{a_1} \right)^{\frac{1}{\alpha-1}}<C_1\leq r$}, \\\\
\frac{(C_1-r)a_1^{\frac{1}{1-\alpha}}}{1+a_1^{\frac{1}{1-\alpha}}} & \hbox{  if \,\,\,$C_1>r$} \\\\
\end{array}
\right.
\end{equation}

\begin{equation}\label{eq:solution_y_OPA}
y^{\mathrm{A}}(C_1,r,a_1)=
\left\{
\begin{array}{lll}
0 & \hbox{  if \,\,\,$C_1\leq r\left(\frac{\lambda}{a_1} \right)^{\frac{1}{\alpha-1}}$} \\\\
C_1-\frac{r-C_1}{(\frac{\lambda}{a_1})^{\frac{1}{1-\alpha}}-1} & \hbox{ if \,\,\,$ r\left(\frac{\lambda}{a_1}\right)^{\frac{1}{\alpha-1}}<C_1\leq r$}. \\\\
r+\frac{C_1-r}{1+a_1^{\frac{1}{1-\alpha}}} & \hbox{  if \,\,\,$C_1>r$} \\\\
\end{array}
\right.
\end{equation}

\subsection{Subproblem B}
Suppose $0<\lambda<a_2$, $0<\alpha<1$
the second subproblem can be written as
\begin{equation}
\begin{array}{lll}
&\underset{x,y}{\max} &a_2x^{\alpha}+U(y;r) \\\\
&\mathrm{s.t.} & x,y\geq 0 \\\\
&&x+y\leq C_2 \\\\
\end{array}
\tag{OP-B}
\end{equation}
\textcolor{black}{where $C_2$ is a given constant.}
Similarly to the first subproblem, the second subproblem can be further simplified as 
\begin{equation}
\begin{array}{lll}
&\underset{x}{\max} & a_2x^{\alpha}+U(C_2-x;r) \\\\
&\mathrm{s.t.} & 0\leq x\leq C_2. \\\\
\end{array}
\end{equation}
For notational convenience we define 
\begin{equation}
f_2(x)=a_2x^{\alpha}+U(C_2-x;r).
\end{equation}
Similarly to OP-A, by studying the first derivative of $f_2(x)$, the solution of OP-B can be derived and expressed for the different possible cases:
\begin{equation}\label{eq:solution_x_OPB}
x^B(C_2)=
\left\{
\begin{array}{lll}
C_2 & \hbox{  if \,\,\,$C_2< r\gamma_1$} \\\\
\frac{(C_2-r)a_2^{\frac{1}{1-\alpha}}}{1+a_2^{\frac{1}{1-\alpha}}} & \hbox{  if \,\,\,$C_2\geq r\gamma_1$} \\\\
\end{array}
\right.
\end{equation}
\begin{equation}\label{eq:solution_y_OPB}
y^B(C_2)=
\left\{
\begin{array}{lll}
0 & \hbox{  if \,\,\,$C_2< r\gamma_1$} \\\\
r+\frac{C_2-r}{1+a_2^{\frac{1}{1-\alpha}}} & \hbox{  if \,\,\,$C_2\geq r\gamma_1$} \\\\
\end{array}
\right.
\end{equation}
where $\gamma_1>1$ and being the unique solution of the following equation:
\begin{equation}
a_2\gamma_1^{\alpha}-\frac{(\gamma_1-1)^{\alpha}}{\left(1+a_2^{\frac{1}{1-\alpha}}\right)^{\alpha-1}}={\lambda}.
\end{equation}

\subsection{Optimal solution of OP-G}
By exploiting the previous results derived for the two auxiliary subproblems, it is possible to fully express the optimal power allocation policy a provider should use to maximize the system social welfare. Denote by $\chi$ the total power budget. Without loss of generality, it is assumed $r_1\leq r_2\leq\dots\leq r_K$, i.e., the values of reference points are in ascending order. Due to the ascending order of reference points, the derivatives of consumers' utilities are in descending order for a given consumption power level $x_1=x_2=\dots=x_K\leq r_1$, that is, $\frac{\partial U(x,r_1)}{\partial x}\geq\frac{\partial U(x,r_2)}{\partial x}\geq\dots\geq\frac{\partial U(x,r_K)}{\partial x} 
(x\leq r_1)$. Since the utility functions are increasing and convex when $x<r_1$, to maximize the sum-utility, it can be easily checked that the first $r_1$ power of the total budget $\chi$ should be allocated to Consumer $1$. After allocating $r_1$ power to Consumer $1$, the utility function of the Consumer $1$ becomes concave and the increasing speed of the utility function slows down. In this situation, it can be noticed that the problem becomes to decide whether continue to allocate power to the  first consumer or start to allocate power to other consumers. Similarly, one can observe that  the utility of consumer $2$ increases more rapidly than other consumers for a  common power $x_2=\dots=x_K\leq r_2$.  Hence, knowing the fact that the first $r_1$ amount of power has been allocated to Consumer $1$, the problem regarding the allocation of the following power $\widehat{C}_2\leq r_2$ can be simplified to decide whether keeping  allocating power to Consumer $1$ or starting to allocate power to Consumer $2$. This problem can be formulated as

\begin{equation}
\begin{array}{lll}
&\underset{x_1,x_2}{\max} &(x_1-r_1)^{\alpha}+U(x_2,r_2) \\\\
&\mathrm{s.t.} & x_1-r_1\geq 0,\,\,x_2\geq 0 \\\\
&&x_1-r_1+x_2\leq \widehat{C}_2 \\\\
\end{array}
\end{equation}
%\textcolor{red}{XXX Li: here it is a little confusing why not using the definition in Eq.(OP-G), thus the sum of U function of all consumers, $U(x_1,r_1)+U(x_2,r_2)$. Also, $C$ is not defined. Is the $C$ the same as $\chi$ in (OP-G)? Why $x_1-r_1+x_2$ instead of $x_1+x_2$?. Also, the following paragraph is hard to understand.}
which can be seen as a special case of subproblem A by setting $x=x_1-r_1$, $y=x_2$ and $C_1=\widehat{C}_2$. According to (\ref{eq:solution_y_OPA}), when $\widehat{C}_2=r_2$, one can obtain $x_2=\widehat{C}_2=r_2$, indicating that the second part of power $r_2$ will be fully allocated to the Consumer $2$. In addition, after firstly allocating $r_1$ to $x_1$ and $r_2$ to $x_2$, the utility function of consumer $1$ coincides with the utility function of Consumer $2$, i.e., $U(x_1=r_1+\Delta,r_1)=U(x_2=r_2+\Delta,r_2)$ for any $\Delta\geq 0$. As a consequence, the following allocation policy to consumer $1$ and $2$ should be the same, namely, no matter how much power will be allocated to the Consumer $1$, the same amount of power should be allocated to the Consumer $2$. Similarly, note that  the utility function of Consumer $3$ increases more rapidly than other consumers for a common power $x_3=\dots=x_K<r_3$, the problem  regarding the third part of the power $\widehat{C}_3\leq r_3$ can be simplified to decide whether continue to allocate power to $x_1$ and $x_2$ or start to allocate power to consumer $3$. This problem can be written as
\begin{equation}
\begin{array}{lll}
&\underset{x_1,x_2,x_3}{\max} &(x_1-r_1)^{\alpha}+(x_2-r_2)^{\alpha}+U(x_3,r_3) \\\\
&\mathrm{s.t.} & x_1-r_1=x_2-r_2\geq 0,\,\,x_3\geq 0 \\\\
&&x_1-r_1+x_2-r_2+x_3\leq \widehat{C}_3 \\\\
\end{array}
\end{equation}
%\Li{The $C$ is the same as in Eq.(19)??? What's the definition?}
Replace $x_1-r_1$, $x_2-r_2$ by $\frac{x}{2}$, and {$x_3$ by $y$}, this problem can be rewritten as:
\begin{equation}
\begin{array}{lll}
&\underset{x,y}{\max} &2^{1-\alpha}x^{\alpha}+U(y,r_3) \\\\
&\mathrm{s.t.} & x,y\geq 0 \\\\
&& x+y\leq \widehat{C}_3 \\\\
\end{array}
\end{equation}
which can be seen as a special case of subproblem A when $2^{1-\alpha}\leq \lambda$. In the rest of the paper, it is assumed that  $(K-1)^{1-\alpha}\leq \lambda$ except otherwise stated, and thus the rest allocation policy for other consumers ($i>3$) can be done in the same manner. OP-G can be  decomposed by a sequence of  subproblems A. Without loss of generality, assume that the power constraint fulfills the following condition (suppose  $r_{K+1}=\infty$):
\begin{equation}
\sum_{i=1}^{J}r_i\leq \chi<\sum_{i=1}^{J+1}r_i, \,\,J\in\{1,\dots,K\}.
\end{equation}
Based on what we have shown before, Consumer $k$ (with $k\in\{1,\dots,J\}$) will be charged at least $r_k$ power and same amount of power beyond $r_k$, i.e., $x_k-r_k=\frac{x_0}{J}\geq 0$, where $x_0=\chi-\sum_{k=1}^J r_k- x_{J+1}$. When $i>J+1$, zero power is allocated to the Consumer $i$. Regarding Consumer $J+1$, its power consumption, $x_{J+1}$, can be obtained jointly with $x_0$ by solving the following problem:
\begin{equation}
\begin{array}{lll}
&\underset{x_0,x_{J+1}}{\max} &J^{1-\alpha}{x_0}^{\alpha}+U(x_{J+1},r_{J+1}) \\\\
&\mathrm{s.t.} & x_0,x_{J+1}\geq 0 \\\\
&& x_0+x_{J+1}\leq \chi-\sum_{i=1}^{J}r_i. \\\\
\end{array}
\label{eq:OP-G}
\end{equation}
%\Li{(XXX Li: for me it is more comfortable to write the optimization problem in its original format, for example here $j\left(\frac{x_0}{j}\right)^{\alpha}$, or even $\sum_j U(x,r)$.)}

\textcolor{black}{Implementing the solution of OP-A, the optimal power allocation policy of OP-G can be written as:
{\footnotesize \begin{equation}\label{eq:solution_x_OPG}
x_i^{\star}(\chi)=
\left\{
\begin{array}{lll}
r_i+\frac{1}{J}x^A(\chi-\sum_{i=1}^{J}r_i,r_{J+1},j^{1-\alpha}) & \hbox{  if \,\,\,$i\leq J$} \\\\
y^A(\chi-\sum_{i=1}^{J}r_i,r_{J+1},J^{1-\alpha})& \hbox{ if \,\,\,$ i=J+1$} \\\\
0 & \hbox{ if \,\,\,$ i>J$} \\\\
\end{array}
\right.
\end{equation}}
where $x^A(.)$ and $y^A(.)$ are defined by (\ref{eq:solution_x_OPA}) and (\ref{eq:solution_y_OPA}), respectively. To maximize the sum-utility, the power will be firstly allocated to consumers with the lowest reference points. While the consumers with the lowest reference points reach their reference point, the system begins to allocate the power to consumers with higher reference points. If the power budget is sufficiently large (e.g., $\chi\geq\sum_{i=1}^K r_i$), the  difference of charged power among the consumers is the same as their reference points difference.}\textcolor{black}{More precisely, by defining the threshold as $T_i$ (assuming $T_{K+1}=\infty$) if and only if the power could be allocated to consumer $i$ when $\mathcal{X}\geq T_i$, the flowchart of the allocation policy can be shown in Fig.~3.}

\begin{figure}[h]
   \begin{center}
        \includegraphics[width=.5\textwidth]{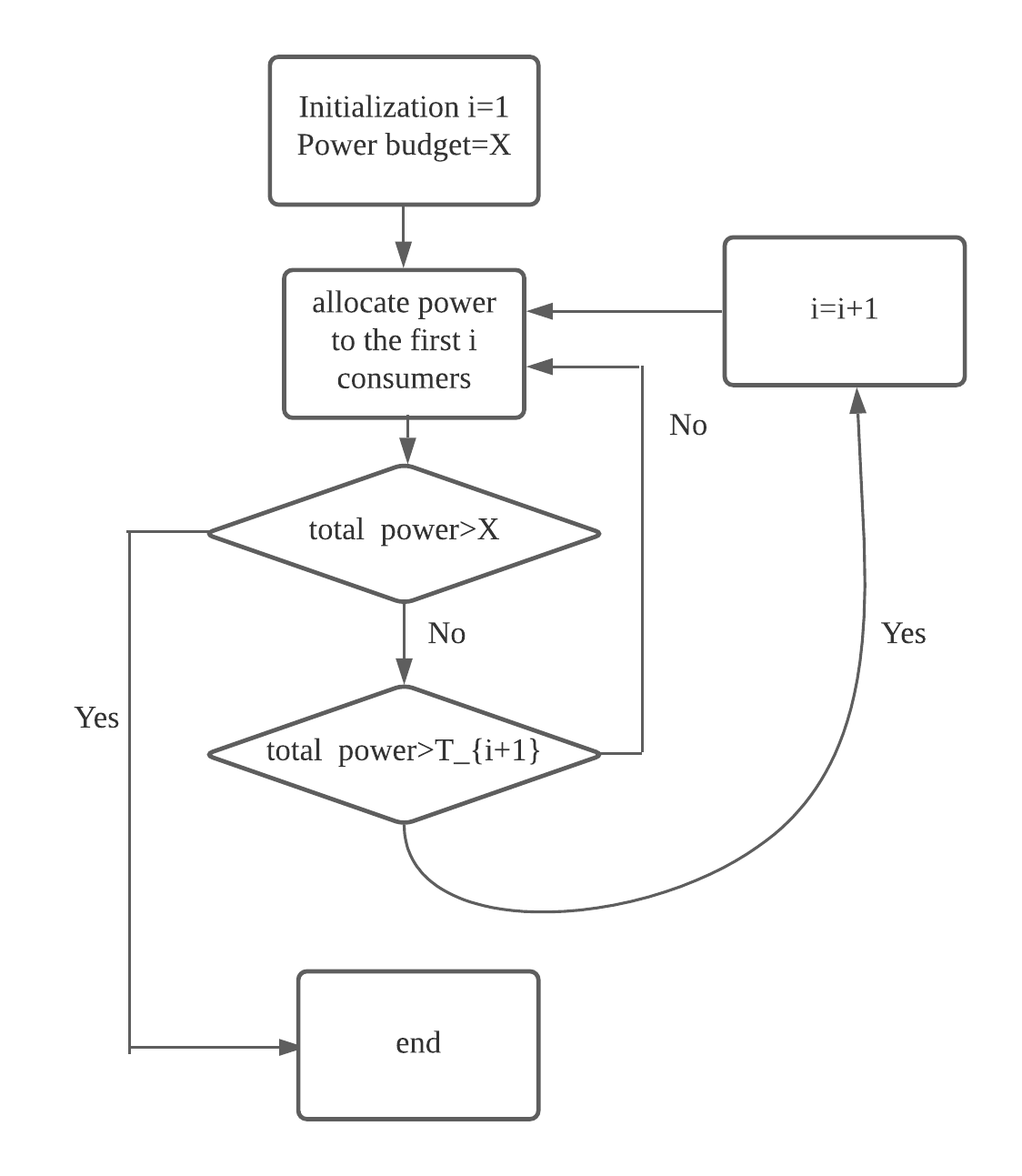}
    \end{center}
        \caption{ \textcolor{black}{Flowchart of the allocation policy}}
\end{figure}

\textbf{Remark:} When $(K-1)^{1-\alpha}>\lambda$, for instance, the subjectivity is not considered and thus $\lambda=1$, the power allocation of (OP-G) can be solved by combining the two subproblems (OP-A) and (OP-B). More precisely, when $k^{1-\alpha}\leq\lambda$ ($k\in\{1,\dots,K\}$), (OP-A) can be firstly used. As $k$ increases to a certain value such that $k^{1-\alpha}>\lambda$, the second subproblem (OP-B) can be applied to solve the problem. For the sake of clarity of the following discussions on pricing policies, we solely consider the scenario where $(K-1)^{1-\alpha}\leq\lambda$. %XXXChao: For this case, it is much more complicated to obtain the expression of the solution as for (\ref{eq:solution_x_OPG}). So I would suggest we solely talk about the case with $(K-1)^{1-\alpha}<\lambda$, at least for analytical results.
\section{Application to the pricing and energy-efficiency problems}
In the previous section, we have considered the problem of maximizing the sum-utility with fixed power consumption. In the perspective of benefits, the optimal total power consumption should also depend on the cost of purchasing (or generating) this amount of power. By using the optimal allocation policy obtained in the preceding section, we consider two more practical scenarios in this section. We consider social welfare (retained gain with cost eliminated) and the global energy-efficiency (benefit brought by unit power consumption) defined by the ratio between the gain modelled by sigmoidal function and the power consumption.
\subsection{Maximization of social welfare with inclining block rates pricing policy}
To maintain a high consumer satisfaction level, we assume that the provider is regulated so that its objective is not to maximize its own profit through electricity trade, but rather to induce users' consumption in a way that maximizes social welfare (see \cite{MR-SGC-2010}\cite{Li-PES-2011}). Social welfare can be seen as the profit of the system, that is, the sum of all consumers' utility minus the cost of providing the electricity demanded by all the consumers:
\begin{equation}
W=\sum_{i=1}^K U(x_i,r_i)-f_c(\chi)
\end{equation}
where $f_c(\chi)$ is the cost function and increasing in $\chi$. For instance, in \cite{MR-SGC-2010}, $f_c(\chi)$ is assumed to be a quadratic function under the form $a\chi^2+b\chi+c$. In  \cite{MR-SGC-2010}, while the utility function is a quadratic concave function w.r.t. $x$, it has been proved that the maximization of social welfare can be achieved by a real-time pricing scheme designed by the provider, with the deployment of demand response. Although the electricity rate of this pricing tariff can vary over time slots, the unit electricity price remains to be constant regardless of the power consumption. However, when pure concavity is no longer available in the model with sigmoidal utility functions, common flat rates are no longer optimal to induce consumer's consumption to maximize social welfare. An alternative to the common flat rates in retail electricity market is the inclining block rates, where the unit rate of electricity changes when the consumption is beyond a certain threshold \cite{MR-TSG-2010}. \textcolor{black}{Intuitively, as the sigmoidal function proposed here has two segments (the first segment with increasing marginal benefit and the second segment with decreasing marginal benefit), the IBR with two price levels and two different linear segments inherently match better our proposed scheme.} Motivated by this intuition, we want to prove that the IBR is the optimal or near optimal pricing tariff to maximize social welfare.

\textcolor{black}{Before introducing the pricing scheme, we first describe the relations between the aforementioned models and the pricing scheme. The sigmoidal function is proposed to model the consumer's benefit. Based on these benefit functions, we study the sum-utility problem aiming to maximize the sum of benefit functions with a fixed overall consumption. The reason to solve  the sum-utility problem is to find an allocation policy that is needed to solve the following social welfare maximization problem defined by (25). After obtaining the optimum consumption profiles to maximize social welfare, the last step is to find a corresponding pricing scheme that could induce consumers' consumption in accordance with the consumption  to maximize social welfare. In the rest of this subsection, we elaborate on the derivation of the pricing scheme.}

As explained before, for a given $\chi$, the optimal power consumption $x_i^{\star}(\chi)$ can be chosen according to (\ref{eq:solution_x_OPG}). Therefore, the maximization of $W$ can be rewritten as follows:
\begin{equation}
\underset{\chi}{\max} \sum_{i=1}^K U(x_i^{\star}(\chi),r_i)-f_c(\chi).
\label{eq:x_sum_op}
\end{equation}
\textcolor{black}{Interestingly, it could be checked that there could have at most one local maximum in the interval $(r_j,r_{j+1})$ for every $j\in\{1,K\}$ (suppose $r_{K+1}=\infty$). Hence we could use conventional approaches (e.g., gradient descent, or Newton method) to find these local maximums and compare them to obtain the global maximum $\mathcal{X}^{\star}$. Or we could use brute-force approach with higher complexity to search the $\mathcal{X}^{\star}$. Since the form of the function $U(x_i^{\star}(\mathcal{X}.,r_i))$ is known, the computational complexity of the single variable brute-force approach is not going to be prohibitive.} More importantly, the value of $\chi^{\star}$ does not change the main structure of the proof of optimality of IBR pricing. For the sake of clarity, suppose the optimal total power $\chi^{\star}$ meets the following condition:
\begin{equation}\sum_{i=1}^{J^{\star}}r_i\leq \chi^{\star}<\sum_{i=1}^{J^{\star}+1}r_i .
\end{equation}
where $J^{\star}\in\{1,\dots,K\}$. As explained in Sec. III-C, for any Consumer $i\in\{1,\dots,J^{\star}\}$, the power consumption beyond its reference point should be the same, namely, $x_1^{\star}(\chi^{\star})-r_1=\dots=x_{J^{\star}}^{\star}(\chi^{\star})-r_{J^{\star}}$. Therefore,  the first derivative of $U(x_i,r_i)$ at optimal power consumption is the same for $1\leq i\leq {J^{\star}}$, that is, 
\begin{equation}
\frac{\partial U(x;r_1)}{\partial x}|_{x=x_1^{\star}(\chi^{\star})}=\dots=\frac{\partial U(x;r_{J^{\star}})}{\partial x}|_{x=x_{J^{\star}}^{\star}(\chi^{\star})}=p
\label{eq:equal_der}
\end{equation}
where $p$ is a constant related to the value of $\chi^{\star}$. Due to the lower benefit brought by the higher values of reference points, the power allocated to any consumers $i\in\{{J^{\star}}+1,\dots,K\}$ is zero. Regarding Consumer ${J^{\star}}+1$, according to (\ref{eq:solution_y_OPA}), if $\chi^{\star}-\sum_{i=1}^{{J^{\star}}}r_i\geq r_{{J^{\star}}+1}(\frac{\lambda}{{J^{\star}}^{1-\alpha}})^{\frac{1}{\alpha-1}}$, the first derivative of $U(x,r_{{J^{\star}}+1})$ at optimal consumption can be checked to be  the same as the first $j$ consumers, i.e., $\frac{\partial U(x,r_{{J^{\star}}+1})}{\partial x}|_{x=x_{{J^{\star}}+1}^{\star}(\chi^{\star})}=p$. Otherwise, if $\chi^{\star}-\sum_{i=1}^{{J^{\star}}}r_i<r_{{J^{\star}}+1}(\frac{\lambda}{{J^{\star}}^{1-\alpha}})^{\frac{1}{\alpha-1}}$, one can obtain $x_{{J^{\star}}+1}^{\star}(\chi^{\star})=x_{{J^{\star}}+2}^{\star}(\chi^{\star})=x_{K}^{\star}(\chi^{\star})=0$. 

%The relation between $p$ and $\chi^{\star}$ can be expressed as:

Assuming the demand response is implemented in the consumer side,  the objective of each consumer is to maximize their own benefit, namely, the individual satisfaction brought by consumption minus the cost of purchasing electricity from the provider, which can be defined as follows:
\begin{equation}
u_i(x_i,r_i)=U(x_i,r_i)-p_i(x_i)
\label{eq:OP_consumer}
\end{equation} 
where $p_i(.)$ represents the cost of user $i$ by consuming $x_i$ amount of power. \textcolor{black}{According to demand response programs, the power consumption is determined by the consumer to maximize their own benefit, so it is significant to guide consumers through tariffs to preserve social welfare.} To maximize social welfare, the provider aims to design appropriate pricing policies such that the optimal power consumption $x_i^{\mathrm{OP}}$ to maximize $u_i(x_i,r_i)$, namely,
\begin{equation}
x_i^{\mathrm{OP}}\in \arg\underset{x_i}{\max}\,\, u_i(x_i,r_i),
\end{equation}
coincides  with $x_i^{\star}(\chi^{\star})$. In the following proposition, we propose an IBR pricing policy such that $x_i^{\mathrm{OP}}=x_i^{\star}(\chi^{\star})$ always holds for all $i\neq {J^{\star}}+1$, and $x_{{J^{\star}}+1}^{\mathrm{OP}}=x_{{J^{\star}}+1}^{\star}(\chi^{\star})$  holds under certain conditions. Hence the IBR pricing policy is optimal or near optimal to induce consumers' consumption such that social welfare can be maximized.
\begin{proposition}
\tc{black}{The optimal power consumption $x_i^{\mathrm{OP}}$ to maximize the individual benefit coincides with the optimal power consumption $x_{i}^{\star}(\chi^{\star})$ to maximize social welfare for all $i\neq {J^{\star}}+1$ by implementing the following pricing policy:
{\footnotesize \begin{equation}\label{eq:pricing}
p_i^{\star}(x_i)=
\left\{
\begin{array}{lll}
qx_i & \hbox{  if \,\,\,$x_i\leq r_i-\left(\frac{p}{\lambda\alpha}\right)^{\frac{1}{\alpha-1}}$} \\\\
(q-p)(r_i-\left(\frac{p}{\lambda\alpha}\right)^{\frac{1}{\alpha-1}})+px_i & \hbox{ if \,\,\,$x_i> r_i-\left(\frac{p}{\lambda\alpha}\right)^{\frac{1}{\alpha-1}}$} \\\\
\end{array}
\right.
\end{equation}}
where 
\begin{equation}
q=\frac{\lambda r_{J^{\star}}^{\alpha}+(\frac{p}{\alpha})^{\frac{\alpha}{\alpha-1}}-((\frac{p}{\alpha})^{\frac{1}{\alpha-1}}+(\frac{p}{\lambda\alpha})^{\frac{1}{\alpha-1}})p}{r_{J^{\star}}-(\frac{p}{\lambda\alpha})^{\frac{1}{\alpha-1}}}+\Delta
\label{eq:q_expression}
\end{equation}
with $\Delta$ can be any negative value lower bounded by 
\begin{equation}
\begin{split}
&-\frac{\lambda r_{J^{\star}}^{\alpha}+(\frac{p}{\alpha})^{\frac{\alpha}{\alpha-1}}-((\frac{p}{\alpha})^{\frac{1}{\alpha-1}}+(\frac{p}{\lambda\alpha})^{\frac{1}{\alpha-1}})p}{r_{J^{\star}}-(\frac{p}{\lambda\alpha})^{\frac{1}{\alpha-1}}}+\\
&\frac{ \lambda r_{{J^{\star}}+1}^{\alpha}+(\frac{p}{\alpha})^{\frac{\alpha}{\alpha-1}}-((\frac{p}{\alpha})^{\frac{1}{\alpha-1}}+(\frac{p}{\lambda\alpha})^{\frac{1}{\alpha-1}})p}{r_{{J^{\star}}+1}-(\frac{p}{\lambda\alpha})^{\frac{1}{\alpha-1}}}.
\end{split}
\label{eq:delta_expression}
\end{equation} 
}
In particular, $x_{{J^{\star}}+1}^{\mathrm{OP}}=x_{{J^{\star}}+1}^{\star}(\chi^{\star})$ can be attained (and hence social welfare maximization can be reconstructed perfectly by the proposed pricing policy) when the following condition is met
\begin{equation}
0<\chi^{\star}-\sum_{i=1}^{{J^{\star}}}r_i\leq r_{{J^{\star}}+1}(\frac{\lambda}{{J^{\star}}^{1-\alpha}})^{\frac{1}{\alpha-1}}
\end{equation}
\end{proposition}
\begin{proof}
See Appendix.
\end{proof}

It is worth mentioning that the proposed pricing scheme is a piecewise function, where the unit price is $q$ before the threshold and becomes $p$ after the threshold. Note that the two unit prices are the same for all the consumers, whereas the threshold of the pricing policy depends on the reference point $r_i$ and thus being different for consumers. To implement the proposed IBR scheme, the provider needs to first find $\chi^{\star}$ as the solution of (\ref{eq:x_sum_op}). Then plugging (\ref{eq:solution_x_OPG}) into (\ref{eq:equal_der}), the constant $p$ can be derived. At last, another rate $q$ can be determined by (\ref{eq:q_expression}) and (\ref{eq:delta_expression}).  Considering the difference between the two price levels, we provide different interpretations for the following three scenarios: $q>p$, $q<p$, and $q=p$. When $q>p$, the optimal pricing policy is to encourage the users to consume large amount of power (discount for large consumption), corresponding to the practical case where the power utility have strong generation capacity which has not been well exploited.  The second case, $q<p$, the optimal pricing policy is to avoid the users to consume large amount of power (punishment for large consumption), corresponding to the practical case where the system have heavy loads and prefer the consumers to cut down their consumption. The last case $q=p$, the proposed IBR pricing policy to maximize social welfare boils down  to constant flat rates pricing policy. This phenomenon can be seen in the cases with special cost functions, e.g., the linear function $f_c(\chi)=a\chi+b$.  

According to Prop. IV.1, for any increasing cost function $f_c(.)$, IBR pricing can at least induce $K-1$ consumers' consumption in the way to maximize social welfare, even though the consumer ${J^{\star}}$ might behave in a different way. Next, with a widely used quadratic cost function, we explore sufficient conditions with which all the consumers' consumption  (including consumer ${J^{\star}}$) coincides with the optimal consumption to maximize social welfare.
\begin{Corollary}
When $C(x)=ax^2+bx+c$, the proposed pricing policy can perfectly reconstruct the optimal power consumption vector to maximize social welfare if the following condition is fulfilled:
\begin{equation}
a\leq\alpha(1-\alpha)r_K^{\alpha-2}.
\end{equation}
\end{Corollary}
\begin{proof}
See Appendix.
\end{proof}

To conclude this part, the proposed IBR pricing policy can be proved to induce  $K-1$ consumers' consumption in a way to maximize social welfare, and all the  consumers will follow the optimal consumption rule under the proposed IBR pricing policy if the cost function satisfies certain conditions. In addition, the proposed pricing tariff is easy to implement in power systems by solving low complexity optimization problems.

\textbf{Remark:} When the minimum need $m_i$ of each consumer is imposed, the problem can be solved in the same way by using an adapted reference point $r_i^{\prime}=r_i-m_i$. The IBR can still be proved to be optimal to induce consumers' behavior in a way to maximize social welfare.

\subsection{Energy-efficiency with sigmoidal utility}

In the preceding section we have shown how the proposed optimization framework can be exploited to maximize social welfare. In this section, we also show that it can be exploited to maximize functions which have a different structure namely, we want to maximize energy-efficiency when it is measured by a ratio being a sum-benefit over a sum-cost (see \cite{Debbah-2016}\cite{Zappone-2016}). The problem can be formulated as
\begin{equation}
\begin{array}{lll}
&\underset{x_1,\dots,x_K}{\max} & \frac{\sum_{i=1}^{K}U(x_i; r_i)}{\sum_{i=1}^{K}x_i} \\\\
&\mathrm{s.t.} & \forall i \in \{1,\dots \}, x_i \geq 0 \\\\
\end{array}
\tag{OP-EE}
\end{equation}
Before deriving the solution of (OP-EE), we firstly introduce some basic definitions. Define $x_i^{\mathrm{EE}}$ as the power consumption to maximize the individual energy-efficiency, that is,
\begin{equation}
x_i^{\mathrm{IEE}}=\arg\underset{x_i > 0}{\max} \,\,\frac{U(x_i ; r_i)}{x_i}
\end{equation}
and define the maximum energy-efficiency can be achieved at consumer $i$  as
\begin{equation}
u_i^{\mathrm{IEE}}=\underset{x_i > 0}{\max} \,\,\frac{U(x_i ; r_i)}{x_i}.
\end{equation}
Note that $U(x_i; r_i)$ is a sigmoidal function, and thus it can be verified that $x_i^{\mathrm{IEE}}>r_i$ is the unique (nonzero) solution of the following equation:
\begin{equation}
x\alpha(x-r_i)^{\alpha-1}-(x-r_i)^{\alpha}-r_i^{\alpha}=0
\label{eq:IEE_solution}
\end{equation} 
The following proposition compares the value of different $u_i^{\mathrm{IEE}}$ and $x_i^{\mathrm{IEE}}$, respectively.
\begin{proposition}
When $1\leq i_2<i_1\leq K$, the following inequalities can be obtained: 
 \begin{equation}
 x_{i_1}^{\mathrm{IEE}}-r_{i_1}> x_{i_2}^{\mathrm{IEE}}-r_{i_2}
 \end{equation}
\begin{equation}
u _{i_1}^{\mathrm{IEE}}< u_{i_2}^{\mathrm{IEE}}
 \end{equation}
 \begin{proof}
See Appendix.
 \end{proof}
\end{proposition}
According to this proposition, one can observe that the maximum individual energy efficiency increases when $r$ decreases, which implies the first consumer (who has the minimum reference points) can achieve the highest individual energy efficiency compared with other consumers. In the following proposition, it can be seen that the solution of (OP-EE) is fully connected to individual energy efficiency solutions.
\begin{proposition}
The solution of (OP-EE), defined as $(x_1^{\mathrm{SEE}},\dots,x_K^{\mathrm{SEE}})$, can be written as
\begin{equation}
x_1^{\mathrm{SEE}}=x_1^{\mathrm{IEE}}
\end{equation}
\begin{equation}
x_i^{\mathrm{SEE}}=0,\,\,\forall i\geq 2
\end{equation}
The maximum sum-energy-efficiency, defined as $u^{\mathrm{SEE}}$, is $ {u_1^{\mathrm{IEE}}}$.
\end{proposition}
\begin{proof}
This proposition can be seen as a special case of Proposition 1 in \cite{Meshkati-2006}. 
%\begin{equation}
%\begin{split}
%&\frac{\sum_{i=1}^{K}U(x_i,r_i)}{\sum_{i=1}^{K}x_i}\leq \frac{\sum_{i=1}^{K}u_i^{\mathrm{IEE}}x_i}{\sum_{i=1}^{K}x_i}
%\leq \max_i {u_i^{\mathrm{IEE}}}
%= {u_1^{\mathrm{IEE}}}
%\end{split}
%\end{equation}
%The equality holds if and only if 
%\begin{equation}
%x_1^{\mathrm{SEE}}=x_1^{\mathrm{IEE}}
%\end{equation}
%\begin{equation}
%x_i^{\mathrm{SEE}}=0,\,\,\forall i\geq 2
%\end{equation}
\end{proof}
Based on this proposition, we see that in order to maximize the energy efficiency of the network, only the first consumer will be charged, which might not be interesting for systems where fairness among users matters. Hence, we consider a more practical scenario in which each consumer obtains its minimum need, $m_i<r_i$, and the objective is to maximize the energy efficiency while satisfying the minimum power need. In addition, the minimum need constraint can be explained from the perspective of the power utility. Shut-down and ramp-up of a power plant can be costly or sometimes not technically feasible. A very low consumption can be detrimental to the power system sustainability. Hence, a minimum supply level can be imposed from the provider as well. The more practical  problem can be formulated as follows:
\begin{equation}
\begin{array}{lll}
&\underset{x_1,\dots,x_K}{\max} & \frac{\sum_{i=1}^{K}U(x_i,r_i)}{\sum_{i=1}^{K}x_i} \\\\
&\mathrm{s.t.} & x_i\geq m_i, \,\,\,\forall 1\leq i\leq K\\\\
\end{array}
\tag{OP-EE-P1}
\end{equation}
Note that $U(x_i,r_i)=U(m_i,r_i)+U(x_i-m_i,r_i-m_i)
$. Replacing $x_i-m_i$ by $\widehat{x}_i$ and $r_i-m_i$ by $\widehat{r}_i$, (OP-EE-P1) can be equivalently written as
\begin{equation}
\begin{array}{lll}
&\underset{\widehat{x}_1,\dots,\widehat{x}_K}{\max} & \frac{M_1+\sum_{i=1}^{K}U(\widehat{x}_i,\widehat{r}_i)}{M_2+\sum_{i=1}^{K}\widehat{x}_i} \\\\
&\mathrm{s.t.} & \widehat{x}_i\geq 0, \,\,\,\forall 1\leq i\leq K\\\\
\end{array}
\tag{OP-EE-P2}
\end{equation}

where $M_1$ and $M_2$ are two constants defined as $M_1=\sum_{i=1}^KU(m_i,r_i)$ and  $M_2=\sum_{i=1}^Km_i$, respectively. Due to the existence of the two constant $M_1$ and $M_2$, $x_i^{\mathrm{SEE}}$ is no longer a solution for (OP-EE-P2). Even though the solution of (OP-EE-P2) cannot be expressed, some properties of (OP-EE-P2) can still be extracted. Suppose
\begin{equation}
E^{\star}=\underset{\widehat{x}_1,\dots,\widehat{x}_K}{\max}  \frac{M_1+\sum_{i=1}^{K}U(\widehat{x}_i,\widehat{r}_i)}{M_2+\sum_{i=1}^{K}\widehat{x}_i}
\label{eq:E_star}
\end{equation}
By studying the first derivative of $U$, one can observe that the power consumption to maximize the EE, defined as $\widehat{x}_i(E^{\star})$, should satisfy the condition either $\frac{U(\widehat{x}_i,\widehat{r}_i)}{\partial \widehat{x}_i}|_{\widehat{x}_i=\widehat{x}_i(E^{\star})}=E^{\star}$ or $\widehat{x}_i=0$. It can be proved that the power consumption can be written as
\begin{equation}\label{eq:EE_solution_constraints}
\widehat{x}_i(E^{\star})=
\left\{
\begin{array}{lll}
0 & \hbox{  if \,\,\,$\frac{\lambda \widehat{r}_i^{\alpha}+(\frac{E^{\star}}{\alpha})^{\frac{\alpha}{\alpha-1}}}{ \widehat{r}_i+(\frac{E^{\star}}{\alpha})^{\frac{1}{\alpha-1}}}\leq E^{\star}$} \\\\
 \widehat{r}_i+(\frac{E^{\star}}{\alpha})^{\frac{1}{\alpha-1}} & \hbox{ otherwise } \\\\
\end{array}
\right.
\end{equation}

Define a function 
\begin{equation}
g(E)= \frac{M_1+\sum_{i=1}^{K}U(\widehat{x}_i(E),\widehat{r}_i)}{M_2+\sum_{i=1}^{K}\widehat{x}_i(E)}-E
\end{equation}
According to (\ref{eq:E_star}), the optimal EE point $E^{\star}$ is a root of the function $g(E)$, i.e.,
\begin{equation}
g(E^{\star})=0
\end{equation}
Moreover, $E^{\star}$ can be proved to be the unique root of $g(E)$ in the following proposition:
\begin{proposition}
There exists a unique $E^{\star}$ such that 
\begin{equation}
g(E^{\star})=0.
\end{equation}
\end{proposition}
\begin{proof}
See Appendix.
\end{proof}
Note  that $\frac{M_1}{M_2}<E^{\star}<u_1^{\mathrm{IEE}}$ since  $g(\frac{M_1}{M_2})>0$ and $g(u_1^{\mathrm{IEE}})<0$. Therefore, to recover the optimal EE point, one can find the roots of $g(E)$ in the interval $(\frac{M_1}{M_2},u_1^{\mathrm{IEE}})$. We resort to numerical approaches  to find $p^{\star}$. As  $g(E)$ has a unique root in the interval $(\frac{M_1}{M_2},u_1^{\mathrm{IEE}})$, the bisection method can be implemented to find the unique root (see  Algorithm.~1). 
%\noindent\rule[0.25\baselineskip]{0.5\textwidth}{1pt}

\begin{algorithm}
%\SetAlgoLined
{\bf{Inputs:}} $\mathrm{ITER}_{\max}$, $\epsilon$\\
{\bf{Outputs:}} $E^{\star}$\\ 
{\bf{Initialization:}} Set iteration index $\mathrm{ITER}=0$. Initialize the $x^{(0)}=\frac{M_1}{M_2}$, $y^{(0)}=u_1^{\mathrm{IEE}}$ and $D=2\epsilon$.  \\
{\bf{While}} {$ D>\epsilon$ and $\mathrm{ITER}<\mathrm{ITER}_{\max}$}{
%{\bf{Outer loop.}} 
\\\quad Calculate the sign of $g(\frac{x^{\mathrm{(ITER)}}+y^{\mathrm{(ITER)}}}{2})$.\\
\quad \quad {\bf{If}} $g(\frac{x^{\mathrm{(ITER)}}+y^{\mathrm{(ITER)}}}{2})\leq 0$ \\\quad \quad\quad$x^{\mathrm{(ITER+1)}}=\frac{x^{\mathrm{(ITER)}}+y^{\mathrm{(ITER)}}}{2}$\\ 
\quad \quad\quad$y^{\mathrm{(ITER+1)}}=y^{\mathrm{(ITER)}}$\\
 \quad \quad {\bf{else}} $y^{\mathrm{(ITER+1)}}=\frac{x^{\mathrm{(ITER)}}+y^{\mathrm{(ITER)}}}{2}$\\ 
 \quad \quad\quad $x^{\mathrm{(ITER+1)}}=x^{\mathrm{(ITER)}}$\\
 \quad\quad{\bf{end If}} \\
 \quad Update the iteration index: $\mathrm{ITER} \gets \mathrm{ITER}+1$.\\
 \quad Update $D=\min (g(|x^{\mathrm{(ITER)}}|),g(|y^{\mathrm{(ITER)}}|)) $.\\
 {\bf{end While}} 
 
%Stop when $r=R$ and go to {\bf{Outer loop}}.
}
$\forall m \in \{1,...,M\}, \,\,\, E^{\star}= g(\frac{x^{\mathrm{(ITER-1)}}+y^{\mathrm{(ITER-1)}}}{2})$

\caption{\small Algorithm to obtain the root of $g(E)$} \label{algo_vector}
\end{algorithm}

%\noindent\rule[0.25\baselineskip]{0.5\textwidth}{1pt}

As a consequence, the optimum power consumption, which maximizes the EE under the minimum need constraints, can be written as
\begin{equation}
x_i^{\mathrm{SEEC}}=m_i+\widehat{x}_i(E^{\star})
\end{equation}
Taking advantage of the uniqueness of the root, the energy-efficiency can be maximized with a low-complexity and fast convergent algorithm.
%XXX: More interpretations on EE.
\section{Numerical analysis}
In this section, we provide simulation results and more precisely assess the performance of the proposed schemes by comparing them with the most relevant existing techniques. For this purpose, the cost function $f_c(x)$ is chosen to be a quadratic function: $f_c(x)=ax^2+bx+c$ with $a=0.05$ and $b=0.5$. Unless stated otherwise, the exponential parameter $\alpha$ is chosen as $\alpha=0.8$ and the loss amplification factor is chosen as $\lambda=1.5$. The system under consideration is deliberately taken to be simple to make the analysis and interpretations easier but, from the computational aspect, the analysis conducted in the paper allows for larger systems to be analyzed. The system consists of $K=5$ consumers connected to a single provider. In the following, we want to compare different power allocation policies and different pricing policies in terms of sum-utility. We start with a power allocation scheme that is sum-utility maximizing, namely, $\max\,\, \sum_i U(x_i,r_i)$.
\subsection{Comparison among different power allocation policies to maximize the sum-utility}
Firstly, we study the problem discussed in Sec. III. The performance comparison is shown for the following three power allocation (PA) strategies: the proposed power allocation policy (optimal), the proportional power allocation policy (PPA), and the uniform power allocation policy (UPA) \cite{Bansal-TWC-2008}. For the PPA, the consumption of Subscriber $i$ scales to its reference point $r_i$, i.e., $x_i=\frac{r_i}{\sum_i r_i}\times \chi$. The reference points vector $\ul{r}=(r_1,\dots,r_K)$ is assumed to be $(1, 1.5,2,2.5,3)$ kW. We select the relative improvement as the metric to assess the performance, that is, the performance difference between our approach and the existing technique over the performance of our method. From Fig.~4, it can be seen that the proposed PA policy can bring up to $30\%$ improvement to the sum-utility. When the total power consumption is less than ${\sum_i r_i}=10$ kW, four peaks can be observed, and the locations of the four peaks are very close to $r_1$, $r_1+r_2$, $r_1+r_2+r_3$ and $r_1+r_2+r_3+r_4$, respectively. This can be explained by the fact that the utility function $U(x_i, r_i)$ changes fast at the point $x_i=r_i$, and thus at the sensitive point $\sum_i^k r_i$ ($k\in \{1,2,3,4\}$), the significance of a good PA will be highlighted. If $\chi$ reaches a certain level and continues to increase, the improvement begins to decrease and becomes negligible, especially compared to the PPA policy.
\begin{figure}[h]
   \begin{center}
        \includegraphics[width=.5\textwidth]{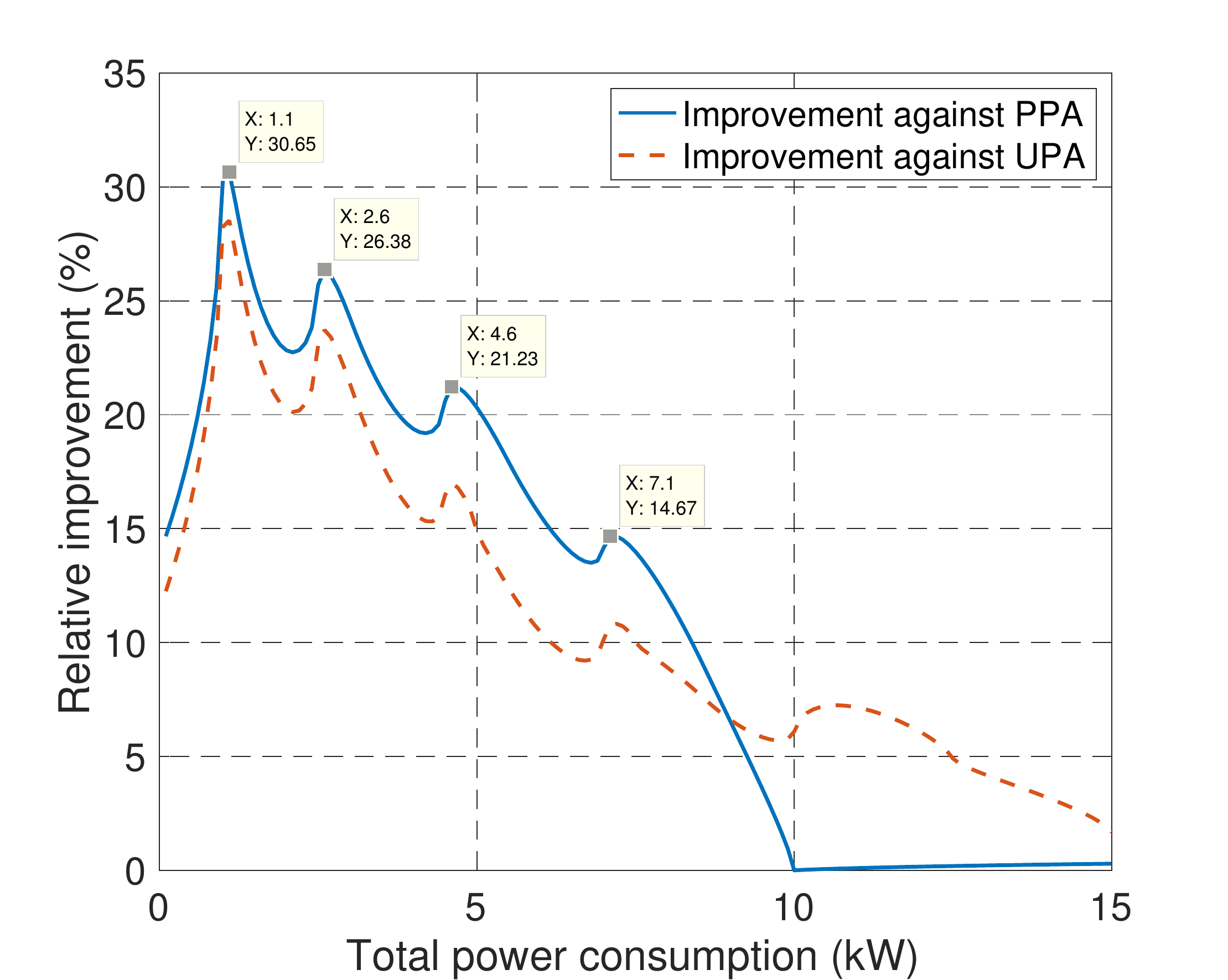}
    \end{center}
    \caption {{Relative utility improvement ($\%$ ) versus sum-utility} for different power allocation (PA) policies. The new PA policy described in Sec. III is shown to perform much better than the conventional PPA and UPA policies, in particular when the power budget becomes relatively small.}
\end{figure}

\begin{figure}[h]
   \begin{center}
        \includegraphics[width=.5\textwidth]{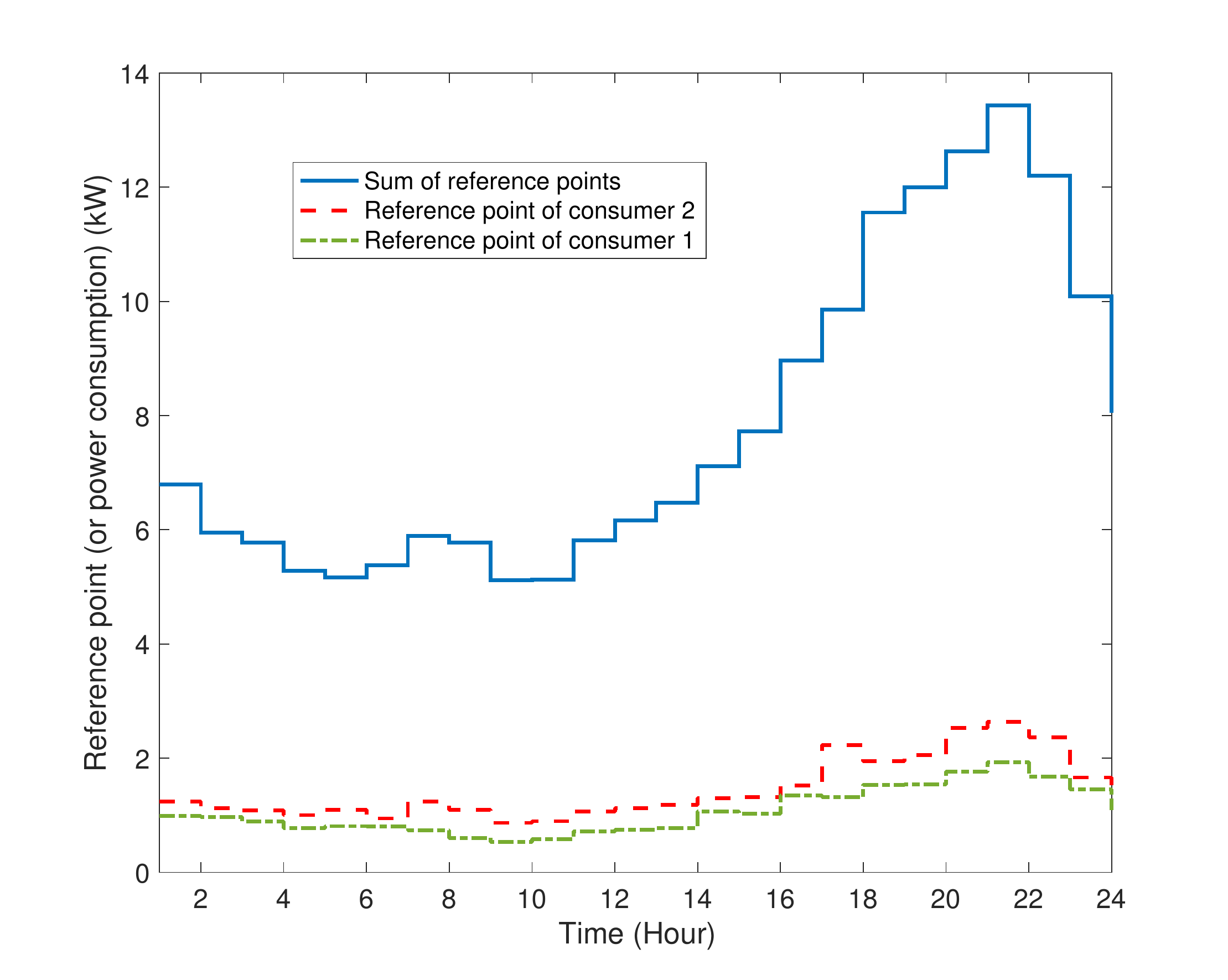}
    \end{center}
        \caption{The reference points for one day by connecting the \textit{PecanStreet} dataset with our model.}
\end{figure}

\begin{figure}[h]
   \begin{center}
        \includegraphics[width=.5\textwidth]{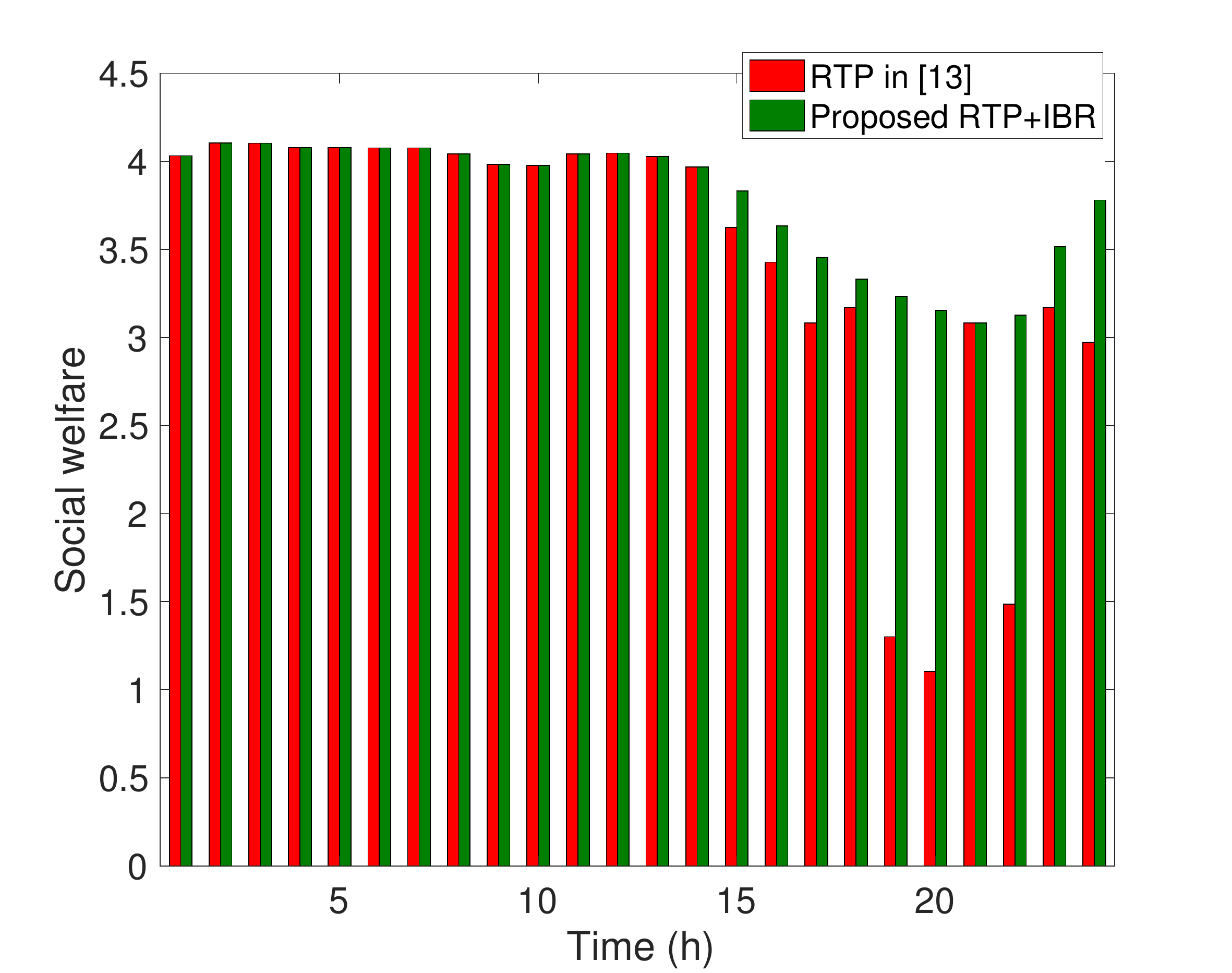}
    \end{center}
        \caption{Social welfare comparison between two pricing tariff policy. The proposed scheme brings a significant improvement when the demand levels are high.}
\end{figure}

\begin{figure}[h]
   \begin{center}
        \includegraphics[width=.5\textwidth]{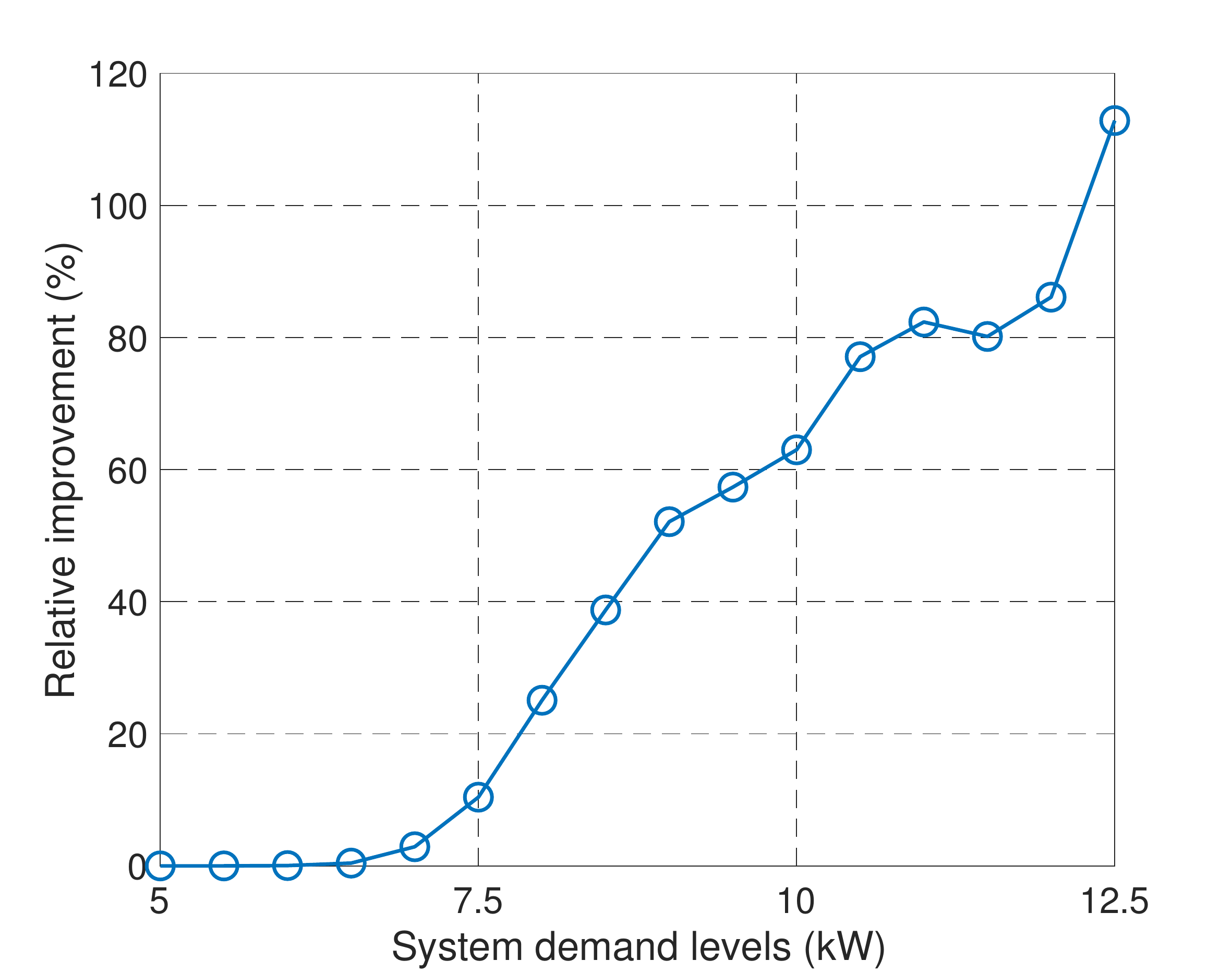}
    \end{center}
        \caption{Relative improvement brought by the proposed scheme against demand levels. The improvement increases rapidly as the system demand levels grows.}
\end{figure}

\begin{figure}[h]
   \begin{center}
        \includegraphics[width=.5\textwidth]{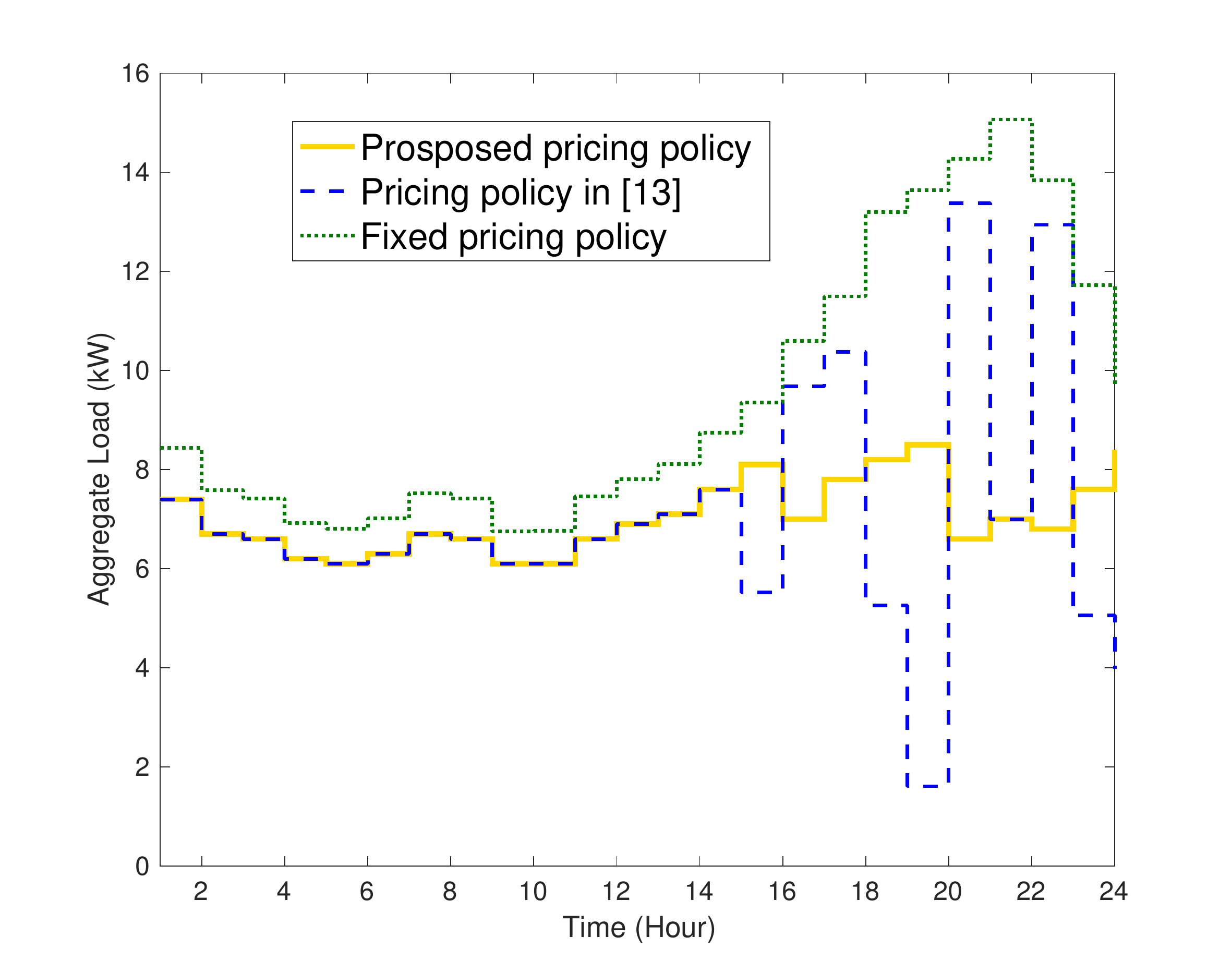}
    \end{center}
        \caption{Aggregate load at different time slots with different pricing policies. Implementing the proposed pricing policy, the aggregate load keeps stable over time, even the demand of different time slots are quite different, which implies that our policy can be a good candidate to minimize the peak power. By contrast, stable power consumption cannot be guaranteed when using state-of-the-art techniques.}
\end{figure}

\begin{figure}[h]
   \begin{center}
        \includegraphics[width=.5\textwidth]{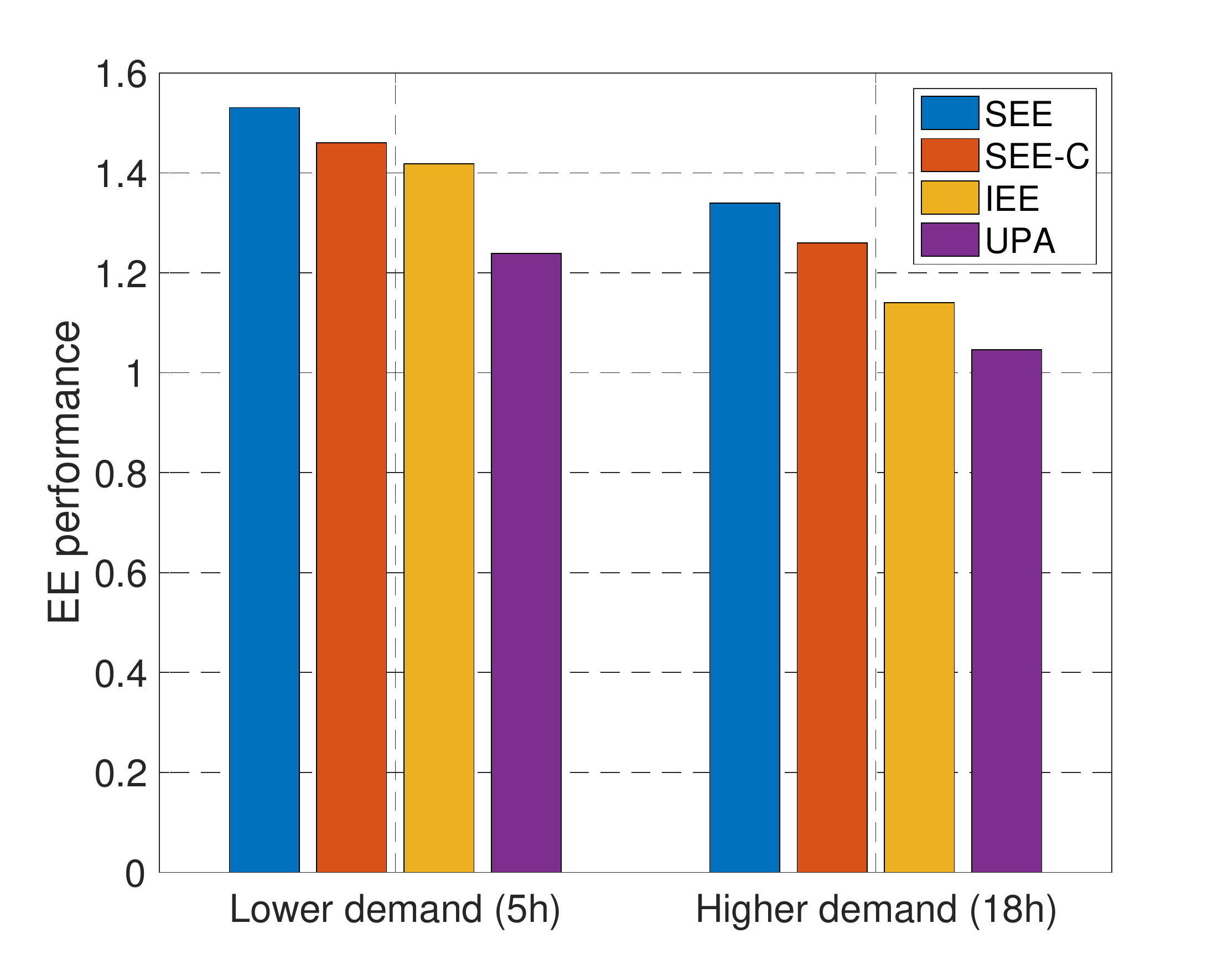}
    \end{center}
       \caption {Sum-energy-efficiency with different techniques. Considering the minImum need constraints, the performance of SEE-C (implemented by Algo.~1) is close to the performance of SEE, which is the maximum performance can be attained without constraints and only the first consumer will be active in SEE policy. Moreover, both SEE and SEE-C can be shown to be more efficient than the classical scheme.}
\end{figure}

%\begin{figure}[h]
%   \begin{center}
%        \includegraphics[width=.5\textwidth]{Simulation/lambda_comparison.eps}
%    \end{center}
%       \caption {Social welfare degradation brought by the wrong prediction over $\lambda$. This degradation is negligible with low demand levels and becomes significant as demand levels increase, which implies that the subjective behavior is not important with low reference points and becomes critical in rush hours. XXXChao: This figure might have some errors, the bar with $\lambda=1$ (underrated) is not very precise.}
%\end{figure}

\subsection{Comparison of pricing policies}

\tc{black}{To quantify the improvement brought by the proposed IBR real-time pricing scheme, we compare it with the real-time pricing scheme proposed in \cite{MR-SGC-2010}, which is independent of consumption and has been proved to be optimal in terms of social welfare for quadratic utility functions. The details about the two pricing policies can be described as follows:
\begin{itemize}
\item{Proposed pricing policy: The IBR pricing tariff is given by (\ref{eq:pricing}). The unit electricity price $q$ in a low consumption regime is given by (\ref{eq:pricing}). Whereas the consumption is beyond a certain threshold, the unit electricity price $p$ is given by (\ref{eq:equal_der}). In each time-slot, due to the different value of reference point (or demand levels), the unit electricity price varies.}
\item{Real-time pricing in \cite{MR-SGC-2010}: The unit electricity price is independent of the consumption and chosen as $p$. The rationale behind this choice is that the marginal benefit of the optimal consumption to maximize social welfare is $p$, and this choice was proven to be optimal in terms of social welfare maximization when the consumers' behaviors are modelled by quadratic utility functions \cite{MR-SGC-2010}. In each time-slot, due to the different value of reference point (or demand levels), the unit electricity price varies.}
\end{itemize} 
To make a reasonable connection between the proposed sigmoidal model with the real data, it is assumed that the reference point is a representation of the past power consumption realizations. We consider five households recorded in \textit{PecanStreet} (Household $26$, Household $4998$, Household $6910$, Household $9499$, Household $9609$) \cite{pecan}\cite{Data-sharing} and use their average power consumption of the Year $2013$ as the reference point of  current time. One day is divided into 24 time slots representing the average power consumption per hour. Fig.~5 shows the sum of reference points(referred to as demand levels) and reference points for some consumers in each hour. The rush hour is in the evening where consumers have higher demand and the least value is achieved at 5 a.m. when most people are still sleeping. In the rest of the simulations, we use this real data as the reference points except otherwise stated. Fig.~6 depicts social welfare with different pricing policies. One can observe that the RTP with constant rates coincides with our proposed RTP with IBR in the morning and early afternoon, but our proposed pricing tariff outperforms the classical RTP in the evening. In fact, it is not necessary to control the consumption when the demand levels are not high, and thus using a well-selected constant rate is sufficient for social welfare maximization. However, as the demand levels increase, an intelligent pricing tariff is required to flatten the consumption and maximize social welfare. With IBR, the consumption behaviors can be optimized by different unit electricity prices. In addition, to see the case in which our proposed scheme can bring more improvement, we conduct simulations to see the improvement with different demand levels. In Fig.~7, we consider a system consisting of five consumers, and the reference point of each consumer is randomly selected from the interval $[\Delta-0.5,\Delta+0.5]$, where $\Delta\in[1,2.5]$. We define the system demand levels as the sum of each individual reference point.  For each given $\Delta$, the average performance is computed over 5000 realizations. To better illustrate the improvement with different demand levels, we define the relative improvement as
\begin{equation}
R_i=\frac{U_{\mathrm{IBR}}-U_{\mathrm{CRTP}}}{U_{\mathrm{CRTP}}}\times 100\%
\end{equation}
where $U_{IBR}$ represents the average social welfare by using our proposed IBR pricing tariff and $U_{CRTP}$ represents the average social welfare by using classical RTP. One can observe that the improvement becomes more significant as the demand levels grows, since our proposed scheme can change the unit electricity price to reduce the load in peak hours.}

\tc{black}{To study the reason why social welfare can be enhanced with our pricing tariff, the aggregate loads at different time slot of the system are illustrated in Fig.~8. Implementing our proposed pricing policy, the aggregate load keeps almost invariant with time, even the demand of different time slots are quite different. This implies that our policy can be a good candidate to minimize the peak power or peak-to-average ratio (PAR). However, with the pricing policy in \cite{MR-SGC-2010}, the aggregate load suddenly increases or decreases in rush hours,  even the aggregate load coincides with our proposed policy while the demand level is not high (from 0h to 15h). The fixed pricing policy chooses the same price for the whole day, and thus the power consumption is proportional to the reference points, resulting in very large peak power.}
\subsection{Sum-energy-efficiency  with different techniques}

The performance of sum-energy-efficiency is assessed for the following four techniques: the sum-energy-efficiency maximization without constraints (SEE),  the sum-energy-efficiency maximization considering constraints (SEE-C), the individual energy efficiency maximization  (IEE) and the UPA. Here, the minimum need is set to be $m_i=\frac{1}{2}r_i$. From Fig.~9, it can be observed that the performance by SEE-C is very close to that of SEE, which means the constraints just bring marginal degradation by using Algo.~1 to find the optimal power vector. Furthermore, both SEE and SEE-C is shown to be better than IEE and UPA, where one aims at maximizing EE of each consumer and another allocate the power uniformly to every consumer. At last, the sum-energy-efficiency decreases when the demand level is higher, for the reason that for higher demands it is more difficult to satisfy the consumer.

\section{Conclusion}
In this paper we propose a refined model for the behavior of an energy consumer. Mathematically, this model consists of a sigmoidal utility function. Moving from the conventional behavior model (namely, a concave utility function) to the proposed refined model implies that some important tasks become more difficult. In particular, the sum-utility (or social welfare) maximization problem becomes more difficult. Although this problem is generally hard computationally speaking, we show that for the considered class of utility functions, the problem can be completely solved and the optimal solutions can be expressed and interpreted. The complete analysis of the considered optimization problem allows one not only to analyze the sum-utility maximization problem but also the problem of a global efficiency whenever measured as the ratio of the sum-benefit to the sum-cost. Solving the sum-utility maximization problem allows us to derive a new pricing policy and more precisely a new inclining block rates policy. The new policy has three attractive features: 1. It is, by construction, optimal in terms of social welfare; 2. It allows the peak-to-average ratio to be managed; 3. In contrast with the conventional real-time pricing policies, the derived pricing policy is not only time-slot dependent but also adapts to the power consumption level. Concerning the energy-efficiency problem, we show that the profit associated with a unit power consumption can be maximized by using a bisection-based algorithm. By constructing a function for which the unique root corresponds to the maximum energy-efficiency, our algorithm is shown to always converge to the global maximum with high convergence speed.

The research work reported in this paper can be extended in many ways. One may better explore the relationship between the reference point and the real energy need, and tune the aversion parameter $\lambda$ accordingly. The parameter selection problem associated with the considered sigmoidal function may be posed and analyzed. To this end, one possible approach is to optimize these parameters by training a deep neural network and using supervised learning. Moreover, the satisfactory level under consideration is modeled by single-stage functions; an interesting and challenging extension might be to use multi-stage utility functions to represent the satisfactory level. \textcolor{black}{Moreover, the model where each consumer has its individual $\lambda$ and $\alpha$ instead of a common $\lambda$ and $\alpha$ can be considered as future works, especially in heterogeneous systems with a great variety of  consumer behaviors.} \textcolor{black}{Finally, the model considering the characteristics of the consumers could be of interest to future works, the approach to the case where consumers have the individual priority might be using hierarchical structure or formulating the problem as a weighted sum-utility problem.}

\section*{Acknowledgement}
This work is partly supported by the RTE-CentraleSupelec Chair, the ANR Project ECOREES and the National Natural Science Foundation of China (No. 51974377).

\section*{Appendix}
\subsection*{A.1. Proof of Prop~4.1}
\begin{proof}
For the piecewise pricing schemes proposed above, it can be easily checked that $x_i^{\mathrm{OP}}\in\{0,r_i+(\frac{p}{\alpha})^{\frac{1}{\alpha-1}}\}$. While the price before the threshold is selected as (\ref{eq:q_expression}), it can be checked that  $x_i^{\mathrm{OP}}=r_i+(\frac{p}{\alpha})^{\frac{1}{\alpha-1}}$ if $i\leq {J^{\star}}$ and $x_i^{\mathrm{OP}}=0$ for $i>{J^{\star}}$. According to (\ref{eq:solution_x_OPA}) and (\ref{eq:solution_y_OPA}), it can be verified that $x_{i}^{\star}(\chi^{\star})=x_i^{\mathrm{OP}}$ if $i\neq {J^{\star}}+1$ and $x_{{J^{\star}}+1}^{\star}(\chi^{\star})$ can be written as
{\tiny\begin{equation}
x_{{J^{\star}}+1}^{\star}(\chi^{\star})=
\left\{
\begin{array}{lll}
0 & \hbox{  if \,\,\,$\chi^{\star}-\sum_{i=1}^{{J^{\star}}}r_i\leq r_{{J^{\star}}+1}(\frac{\lambda}{{J^{\star}}^{1-\alpha}})^{\frac{1}{\alpha-1}}$} \\\\
r_{{J^{\star}}+1}-(\frac{p}{\lambda\alpha}^{\frac{1}{\alpha-1}}) & \hbox{ if \,\,\,$ \chi^{\star}-\sum_{i=1}^{{J^{\star}}}r_i>r_{{J^{\star}}+1}(\frac{\lambda}{{J^{\star}}^{1-\alpha}})^{\frac{1}{\alpha-1}}$} \\\\
\end{array}
\right.
\end{equation}}
When the first condition is met, $x_{{J^{\star}}+1}^{\mathrm{OP}}$ coincides with $x_{{J^{\star}}+1}^{\star}(\chi^{\star})$. Therefore, by implementing the proposed pricing policy, at least the power consumption of $K-1$ consumers  coincides with the power consumption to maximize social welfare (${J^{\star}}+1$)-th consumer is the one which might have different consumption with the targeted consumption to maximize social welfare, and the pricing policy can perfectly reconstruct the every $x_i^{\star}(\chi^{\star})$ when the following condition is fulfilled:
\begin{equation}
0<\chi^{\star}-\sum_{i=1}^{{J^{\star}}}r_i\leq r_{{J^{\star}}+1}(\frac{\lambda}{{J^{\star}}^{1-\alpha}})^{\frac{1}{\alpha-1}}
\end{equation}
\end{proof}

\subsection*{A.2. Proof of Corollary 1}
\begin{proof}
Define $H(x)=\sum_{i=1}^KU(x_i^{\star}(x),r_i)$ and its derivative as 
\begin{equation}
h(x)=\frac{\partial H(x)}{\partial x}, \quad x\neq \sum_{i=1}^k r_i,\,\forall 1\leq k\leq K
\end{equation}
It can be checked that $h(x)$ is partially convex in the intervals $(0,r_1)$, $(\sum_{i=1}^{k}r_{i},\sum_{i=1}^{k+1}r_{i})$, respectively, where $1\leq k\leq K$ (not globally convex in the whole interval $(0,\sum_{i=1}^Kr_{i})$). In the interval $(\sum_{i=1}^{{J^{\star}}}r_{i},\sum_{i=1}^{{J^{\star}}+1}r_{i})$, h(x) firstly decreases until the point $r_{J^{\star}}^{\star}=\sum_{i=1}^{{J^{\star}}}r_{i}+(\frac{\lambda}{{J^{\star}}^{1-\alpha}})^{\frac{1}{\alpha-1}}r_{{J^{\star}}+1}$, then becomes increasing when $x>r_{J^{\star}}^{\star}$. The solution of $\chi^{\star}$ should satisfy $h(\chi^{\star})=C'(\chi^{\star})$, thus we study the position of the intersections between $h(x)$ and $C'(x)$. Two scenarios are possible: both intersections are larger than $r_{J^{\star}}^{\star}$ (case I), or one is larger than $r_{J^{\star}}^{\star}$ and another is smaller than $r_{J^{\star}}^{\star}$ (case II).
%\begin{figure}[H]
%\begin{tabular}{cc}
%\includegraphics[width=.2\textwidth]{Simulation/proof_bad.eps} & \includegraphics[width=.2\textwidth]{Simulation/proof_good.eps} \\
%I & II 
%\end{tabular}
%\end{figure}

For both cases, one can check that $h(x)$ is larger than $C'(x)$ when $x$ is smaller than the left intersection or bigger than the right intersection. Thus, the $\chi^{\star}$ can be solely the left intersection. According to Proposition IV.1, if the left intersection point is always less than $r_j^{\star}$, the perfect reconstruction can be attained by the IBR pricing policy. Therefore, it boils down to finding the sufficient condition such that only case II can happen. Note that $h(x)$ is not differentiable at $x=r_i^{\star}$, and thus there exists a minimum $|h'(x)|>0$. It can be checked that, if the second derivative of $C(x)$, i.e., $C''(x)$, is lower than the   
 minimum second derivative of $H(x)$, i.e., $h'(x)$, then the occurrence of case I can be always averted. By some derivations, it can be demonstrated that
 \begin{equation}
 h'(x)>\alpha(1-\alpha)r_K^{\alpha-2}
 \end{equation}
 Consequently, our claim is proved.
\end{proof}
\subsection*{A.3. Proof of Prop~4.2}
\begin{proof}
 According to (\ref{eq:IEE_solution}), we can write 
\begin{equation}
x_{i_1}^{\mathrm{IEE}}\alpha(x_{i_1}^{\mathrm{IEE}}-r_{i_1})^{\alpha-1}-(x_{i_1}^{\mathrm{IEE}}-r_{i_1})^{\alpha}-r_{i_1}^{\alpha}=0
\label{eq:proof_IEE1}
\end{equation} 
\begin{equation}
x_{i_2}^{\mathrm{IEE}}\alpha(x_{i_2}^{\mathrm{IEE}}-r_{i_2})^{\alpha-1}-(x_{i_2}^{\mathrm{IEE}}-r_{i_2})^{\alpha}-r_{i_2}^{\alpha}=0
\label{eq:proof_IEE1}
\end{equation} 
Assume $x=x_{i_2}^{\mathrm{IEE}}+r_{i_1}-r_{i_2}$, we have 
\begin{equation}
\begin{split}
&x\alpha(x-r_{i_1})^{\alpha-1}-(x-r_{i_1})^{\alpha}-r_{i_1}^{\alpha}\\
=&(x_{i_2}^{\mathrm{IEE}}+r_{i_1}-r_{i_2})\alpha(x_{i_2}^{\mathrm{IEE}}-r_{i_2})^{\alpha-1}-(x_{i_2}^{\mathrm{IEE}}-r_{i_2})^{\alpha}-r_{i_1}^{\alpha}\\
=&(r_{i_1}-r_{i_2})\alpha(x_{i_2}^{\mathrm{IEE}}-r_{i_2})^{\alpha-1}+r_{i_2}^{\alpha}-r_{i_1}^{\alpha}\\
>&(r_{i_1}-r_{i_2})\alpha(x_{i_2}^{\mathrm{IEE}}-r_{i_2})^{\alpha-1}-(r_{i_1}-r_{i_2})(\alpha r_{i_2}^{\alpha-1})\\
=&\alpha(r_{i_1}-r_{i_2})((x_{i_2}^{\mathrm{IEE}}-r_{i_2})^{\alpha-1}-r_{i_2}^{\alpha-1}).
\end{split}
\end{equation}
Note that 
\begin{equation}
2r_{i_2}\alpha(2r_{i_2}-r_{i_2})^{\alpha-1}-(2r_{i_2}-r_{i_2})^{\alpha}-r_{i_2}^{\alpha}<0
\end{equation}
Therefore, we can obtain $r_{i_2}<x_{i_2}^{\mathrm{IEE}}<2r_{i_2}$.  Consequently, it can be checked that 
\begin{equation}
x\alpha(x-r_{i_1})^{\alpha-1}-(x-r_{i_1})^{\alpha}-r_{i_1}^{\alpha}>0
\end{equation}
which implies that 
\begin{equation}
x_{i_1}^{\mathrm{IEE}}-r_{i_1}>x_{i_2}^{\mathrm{IEE}}-r_{i_2}
\end{equation}
According to the definition of U in (\ref{eq:utility_simplified}), one can easily get 
\begin{equation}
u _{i_1}^{\mathrm{IEE}}< u_{i_2}^{\mathrm{IEE}}
\end{equation}
\end{proof}
\subsection*{A.4. Proof of Prop~4.4}
\begin{proof}
Here, we use the proof by contradiction. Suppose there exists $E_1$ and $E_2$ ($E_1>E_2$) such that $g(E_1)=g(E_2)=0$. According to (\ref{eq:EE_solution_constraints}) and the properties of sigmoidal function, it can be verified that if $\frac{\lambda \widehat{r}_i^{^\alpha}+(\frac{E_1}{\alpha})^{\frac{\alpha}{\alpha-1}}}{  \widehat{r}_i+(\frac{E_1}{\alpha})^{\frac{1}{\alpha-1}}}\leq E_1$,
\begin{equation}
U(\widehat{x}_i(E_2);\widehat{r}_i)-U(\widehat{x}_i(E_1),\widehat{r}_i)=0.
\end{equation}
Otherwise, we have
\begin{equation}
U(\widehat{x}_i(E_2),\widehat{r}_i)-U(\widehat{x}_i(E_1),\widehat{r}_i)\geq E_2(\widehat{x}_i(E_2)-\widehat{x}_i(E_1)).
\end{equation}
Consequently, one can obtain that
\begin{equation}\label{eq:frac1_proof}
\frac{\sum_{i=1}^{K}(U(\widehat{x}_i(E_2),\widehat{r}_i)-U(\widehat{x}_i(E_1),\widehat{r}_i))}{\sum_{i=1}^{K}(\widehat{x}_i(E_2)-\widehat{x}_i(E_1))}\geq E_2
\end{equation}
Furthermore, note that 
\begin{equation}\label{eq:frac2_proof}
\frac{{M_1+}\sum_{i=1}^{K}U(\widehat{x}_i(E_1),\widehat{r}_i)}{{M_2+}\sum_{i=1}^{K}(\widehat{x}_i(E_1))}=E_1>E_2.
\end{equation} 
Hence the sum of the two fractions defined in (\ref{eq:frac1_proof}) and (\ref{eq:frac2_proof}) (sum over denominator and nominator respectively) should be larger than $E_2$, i.e., 
\begin{equation}
\frac{{M_1+}\sum_{i=1}^{K}U(\widehat{x}_i(E_2),\widehat{r}_i)}{{M_2+}\sum_{i=1}^{K}(\widehat{x}_i(E_2))}>E_2
\end{equation}
This implies
\begin{equation}
g(E_2)>E_2.
\end{equation}
which leads to the contradiction. Therefore, there exist a unique root of $g(E)$.
\end{proof}

\section*{Reference}
\bibliographystyle{IEEEbib}
\bibliography{strings,refs}

\end{document}